\documentstyle[galley,graphicx,epsf]{mn2e}

\begin{document}

\title {Elemental Abundance Survey of The Galactic Thick Disk}

\author[ Reddy et al.]
        { Bacham E. Reddy$^{1}$, David L. Lambert$^{2}$, Carlos Allende Prieto$^{2}$\\
$^{1}$  Indian Institute of Astrophysics, Bangalore 560034, India\\
$^{2}$  The W.J. McDonald Observatory, University of Texas, Austin, Texas 78712, USA}

\maketitle

\label{firstpage}

\begin{abstract}
We have performed an abundance analysis
 for F- and G- dwarfs of the Galactic thick disk component.
A sample of 176 nearby ($d \leq 150$ pc) thick disk candidate
stars
was chosen from the $Hipparcos$ catalogue and subjected
to a high-resolution spectroscopic analysis.
Using accurate radial velocities combined with  $Hipparcos$ astrometry,
kinematics $(U, V,$ and $~W)$ and Galactic orbital
parameters
were computed. We estimate the probability for a star to belong to
the thin disk,  the thick disk or the halo.
With a probability $P \geq 70$\%  taken as certain membership,
we assigned 95 stars to the thick disk,
13 to the thin disk, and 20 to the halo. The remaining 48 stars in the sample
cannot be assigned with reasonable certainty to one of the three components.

Abundances of C, O, Na, Mg, Al, Si, Ca, Sc, Ti,
V, Cr, Mn, Fe, Co, Ni, Cu, Zn,
Y, Ba, Ce, Nd, and Eu have been obtained. The abundances for thick disk stars
are compared with those for thin disk members
from Reddy et al. (2003). The ratios of $\alpha$-elements (O, Mg, Si, Ca and Ti)
to iron for thick disk disk stars
show a clear enhancement compared to thin disk members in the range
$-0.3 <$ [Fe/H] $ < -1.2$. There are also other
elements --  Al, Sc, V, Co, and possibly  Zn -- which show enhanced ratios
to iron in the thick disk relative to the thin disk.
The abundances of Na, Cr, Mn, Ni,
and Cu (relative to Fe) are very similar for thin and thick disk stars.
The dispersion in abundance ratios [X/Fe] at given [Fe/H] for thick disk stars
is consistent with the expected scatter due to measurement errors, suggesting
a lack of `cosmic' scatter.

A few stars classified as  members of the thick disk by our kinematic criteria
show thin disk abundances.
These stars, which appear older than most thin disk stars, are also, on average,
younger than the thick disk population. They may have originated early
in the thin disk history, and been subsequently scattered to hotter orbits
by collisions. The thick disk may not include stars with [Fe/H] $> -0.3$.
The observed compositions of the thin and thick disks seem to be consistent
with models of galaxy formation by hierarchical clustering
in a $\Lambda$CDM universe. In particular, the
distinct abundance  patterns observed in the thin and thick disks, and the
chemical homogeneity of the thick disk at different galactocentric
distances favor a scenario in which the majority of thick-disk stars
were formed {\it in situ}, from gas rich merging blocks.

\end{abstract}

\begin{keywords}
stars: atmospheric parameters-- stars: abundances -- stars: thick and thin disc --
stars: kinematics -- Galaxy: evolution -- Galaxy: abundances

\end{keywords}

\section{Introduction}
Stars of the solar neighbourhood are overwhelmingly members of the
Galactic disk, with a small admixture of halo stars. The assignment of a
star to
the disk or the  halo is based on  differences in chemical composition and
kinematics.  The local disk population is subdivided into
stars of the thin disk and others belonging to the thick disk, with
chemical composition and kinematics again playing a role in effecting this
subdivision.

The modern division of the disk into  the thin and
thick disk was proposed by Gilmore \& Reid (1983).
 Star counts
led them to divide the disk population in the solar neighbourhood into a thin disk with a
scale height of 300 pc and a thick disk with the much greater
scale height of 1450 pc. Thin disk stars
outnumber thick disk stars by about twenty to one  in the Galactic
plane.
Many other estimates of scale heights and
relative densities of thin and thick disk populations
now exist (e.g., Buser et al. 1999; Ojha 2001;
Cabrera-Lavers, Garz\'on \& Hammersley 2005; Juric et al. 2005).
The thick disk stars are generally older than
most thin disk stars. The metallicity distribution of the thick
disk population is shifted to lower values relative to the
distribution for the thin disk
by about 0.5 dex. Although both distributions can be reasonably
approximated by Gaussians with a FWHM of roughly 0.5 dex, the thick disk
includes a tail at lower metallicities.
In contrast to the thin disk stars, which orbit the Galactic
centre on nearly circular orbits, the thick disk stars are on
moderately elliptical orbits
that typically reach higher distances from the plane. Thick
disk stars also revolve around the Galactic center
slower than those in the thin disk.

The origin of the thick disk has
occasioned much debate.
Keys to the origin lie within the kinematics and the composition
of the thick disk stars.
A number of recent
spectroscopic studies have set out to compare the
chemical compositions of thick and thin disk stars. This
avenue was explored by Gratton et al. (1996) and Fuhrmann (1998).
Gratton et al. showed that O/Fe ratios for thick disk stars
are distinctly different from thin disk stars but similar to halo stars.
Fuhrmann confirmed this based on Mg abundances and
showed a clear cut difference in the Mg/Fe ratio for thick and
thin F-G dwarf stars of the same [Fe/H]\footnote{Here and throughout
the paper we use the so-called {\it bracket} notation to indicate chemical
abundance ratios of two elements X and Y:
[X/Y]  $\equiv \log$ N(X)/N(Y) $- \log $ (N(X)/N(Y))$_{\odot}$}.
These studies stimulated
several investigations of elemental abundances in samples
of thick and thin disk stars -- see, for example, Prochaska et al. (2000),
Feltzing, Bensby,
\& Lundstr\"{o}m (2003), Reddy et al. (2003, hereafter Paper I),
Bensby, Feltzing, \& Lundstr\"{o}m (2003, 2004),
Bensby et al. (2005), and Mishenina et al. (2004).
Although the pattern of
abundance differences between thick and thin disk is emerging, many
details remain obscure, largely, one suspects, because these
investigations cover small numbers of thick disk stars: Prochaska et al.
considered ten, Bensby and colleagues analyzed 36, and Mishenina et al.
less than 30 stars. Considering that the thick disk may
span a range of 1 dex in [Fe/H], these samples, even when
combined, are probably too small to define in detail the differences
between  compositions of  thick and thin disk stars over their
full range in [Fe/H], even if the two disk components were
themselves  chemically homogeneous as a function of
metallicity. Additionally, different definitions of what constitutes
a thick disk star have been adopted by different authors.

Tens of thousands of thick-disk stars at a few kpc from the plane
have been spectroscopically observed as part of the
Sloan Digital Sky Survey
(York et al. 2000; Adelman-McCarthy et al. 2005). These spectra, however,
have a much lower resolving power than the surveys mentioned above, and
although they may yield abundance ratios for a number of metals that
produce strong spectral lines, they have, to this date,
been used to derive iron abundances only (Allende Prieto et al. 2005).

Exploration of the chemical compositions of local thin disk stars
is now well advanced. In particular, several surveys have
investigated
many elements in F-G dwarfs whose spectra are amenable to
quantitative analysis. Our recent study of 26 elements in
181 F-G dwarfs (Paper~I) was the precursor for the
work presented in this paper. The vast majority of the 181
stars belong to the thin disk, as judged (see below) by their
kinematics. Our Paper~I sample may be combined with other
large samples to which thin disk stars are the major
contributor: e.g., Edvardsson et al. (1993) for 189 stars,
and Chen et al. (2000) for 90 stars.  A key result of
our 2003 survey was the finding that `cosmic' scatter in
an abundance ratio X/Fe at a given Fe/H for thin disk
stars was less than the
small measurement errors. Here, we  apply the
same analytical techniques to a large sample of thick stars
for which the cosmic scatter and, indeed, the form of the
run of [X/Fe] with [Fe/H] was not known at the outset of this
project. There were clear indications of the sign and magnitude
of some abundance differences between thick and thin disk,
as recognized by Fuhrmann (1998), and Prochaska et al. (2000) and
further examined by Bensby and colleagues, and by Mishenina et al.
(2004).

The present survey provides abundances for 23 elements
from C to Eu for  176 stars in the solar neighbourhood,
of which 95 are attributed to the thick disk.
The full sample
is introduced in the next section. The observations and abundance analysis
are based closely on Paper~I's approach is described
in Section 3 and 4. Full results, and comparisons with other studies
are given in Section 5. Chemical evolution of
the thick disk and evolution of the Galactic disk are
discussed in Section 6. This section includes
discussion on stars which have thick disk kinematics and
thin disk abundances (TKTA), disk heating, and
merger scenarios. Concluding remarks are  given in Section 7.

\section{The thick disk sample}

\subsection{Preliminary selection}

Stars were first selected from the {\it Hipparcos} catalogue
according to the following criteria: a declination north of
$-30^\circ$ so that they were observable from the
W. J. McDonald Observatory; a B$-$V colour corresponding to
an effective temperature of 5000 -- 6500 K which eliminates
the cool dwarfs with rich line spectra and the hotter dwarfs
where rapid rotation may broaden lines;  and an absolute visual
magnitude in the range $2.5 \leq M_{\rm V} \leq 6.0$
indicating evolution off the zero-age main
sequence so that an age determination is possible; a distance
of less than 150 pc to avoid significant uncertainties in the
observed trigonometric parallaxes, and the introduction of a reddening correction.
Application of these criteria provided about 9,300 stars from the
catalogue's total of 118,218.

The {\it Hipparcos} catalogue was the sole source of
parallaxes and  proper motions. Our initial selection of
thick disk candidates  was based on space motions computed using
radial velocities collected from
several catalogues: the {\it Hipparcos} Input Catalogue,
Carney et al. (1994), Barbier-Brossat \& Figon (2000), and
Malaroda, Levato, \& Galliani  (2001). These sources provided radial velocity
data for 1,970 of the 9,300 stars.
The space velocities $U,V,W$ were computed, where $U$ is directed towards the
Galactic centre, $V$ is directed in the sense of Galactic rotation
(clockwise as seen from the North Galactic Pole), and $W$
is directed towards the North Galactic Pole. Then, the velocities
$U_{LSR}, V_{LSR},
W_{LSR}$ relative to the local standard of rest  were
calculated  assuming the
solar motion of
($U,V,W$) $=$ $(10.0,5.3,7.2)$ km s$^{-1}$, as
derived by Dehnen \& Binney (1998) from {\it Hipparcos} data.
After calculation of their
$U_{LSR}$, $V_{LSR}$, and $W_{LSR}$, a selection by $V_{LSR}$ and $W_{LSR}$
was made to increase the yield of
thick disk (and halo)
stars: $V_{LSR} \leq -40$ km s$^{-1}$ and $\left|W_{LSR}\right|
\geq 30$ km s$^{-1}$.
These criteria led to a sample of 213 stars which was reduced to
176
when inspection of spectra
revealed stars with broad lines and the double-lined spectroscopic
binaries.

In the final computation of the $U,V,W$ for the 176 stars, we made
extensive use of three new sources of radial velocities:
Latham et al. (2002), Nidever et al. (2002), and Nordstr\"{o}m et al. (2004).
The agreement between these sources for stars in common
corresponds to about $\sigma = 0.2$ km s$^{-1}$.
For 146 of the 176  stars, the radial velocity is given in one or
more of these sources. For the remaining 30 stars, we
derived the radial velocity from our spectra with an accuracy of
about 0.5 km s$^{-1}$.

Space velocities were recomputed using the new radial velocities.
We examined the effects of errors in the
input parameters (parallax, proper motion, and
radial velocity), assuming these errors are uncorrelated. The
uncertainties in $U_{LSR}$, $V_{LSR}$, and $W_{LSR}$ are calculated as the quadratic
sum of the individual uncertainties in the input parameters.
In a typical case, the Galactic velocity components are
accurate to about 5 km s$^{-1}$. An uncertain parallax
(say, 30\% accuracy) can lead to a much larger uncertainty; there are
12 stars for which the uncertainty in one of the component velocities is as
large as 30-50 km s$^{-1}$.

\renewcommand{\thetable}{1}
\begin{table}
\centering
\caption{Given are the  velocity dispersions,
the asymmetric drift velocities relative to local standard of rest ($V_{ad}$), and the fractional population of
three stellar components: thin, thick and the halo.
}
\begin{tabular}{@{}lrrrrr@{}}
\hline \hline
Component   & $\sigma_{U}$ & $\sigma_{V}$ &$\sigma_{W}$& V$_{ad}$ & Fraction  \\
\hline
Thin disk    &     43     &     28     &  17      &  $-$9 &0.93   \\
Thick disk   &      67      &       51     &   42       &  $-$48   & 0.07\\
Halo         &      131     &      106     &   85       &   $-$220 & 0.006\\
\hline
\end{tabular}
\end{table}

\subsection{Membership probabilities}

In developing an understanding of the differences between the thin
and thick disks, an obvious prerequisite is a reliable method, even if
statistical,  of assigning a star to the thin or thick
disk, and of  recognizing stars for which an assignment cannot be made
with fair certainty. Since our goal is to quantify the differences in chemical
composition between the thin and thick disk, composition cannot
be a consideration in determining membership. Kinematic criteria
are invoked for this purpose. We follow recent studies in
developing and applying the criteria. The criteria are blunt
instruments; the kinematics of especially thin disk stars but one
suspects also of thick disk stars are complicated.

The thin disk, as represented by the F-G dwarfs in the
solar vicinity, is neither a  monolithic structure
nor cleanly separated from the thick disk
in
$U_{LSR}$, $V_{LSR}$, $W_{LSR}$ space.
Structures include  moving groups or stellar streams
which are seen as regions of enhanced stellar density
in  phase space (see Nordstr\"{o}m et al. 2004; Famaey et al. 2005).
In addition, the mean location of stars
and the dispersion about their mean velocity is dependent on the
age of the stars.

Characteristics  of thick disk stars are the generally  negative
$V_{LSR}$ and the larger $W_{LSR}$ (relative to the thin
disk).
At present, thick disk stars with accurate
$U_{LSR}$, $V_{LSR}$, and $W_{LSR}$  velocities  are too few
in number to identify moving groups should they exist.
It is commonly agreed that the dispersions in the
$U_{LSR}$, $V_{LSR}$ and $W_{LSR}$ velocities
of the thick disk stars exceed those of the thin disk stars.

There are regions in $U_{LSR}$, $V_{LSR}$, $W_{LSR}$ space where
both thin and thick disk stars may occur.
In such boundary layers
shared by thin and thick stars,
one must resort
to a probability argument.  Boundary layers also exist
mixing thick disk with halo stars and again a probability
argument must be used.

In this paper, the method of assigning the probability for each star to either
the thin disk, the thick disk, or the halo is basically that adopted in
the earlier studies by Bensby et al. (2003, 2004) and Mishenina et al. (2004).
We assume the  sample is a mixture of the three
populations.
These populations are each
assumed to be represented by Gaussian distribution
functions for the velocity components $U_{LSR}$, $V_{LSR}$, and $W_{LSR}$,
with given  mean values and dispersions.
(The well-known age dependence of the quantities for
the thin disk is ignored.)  The remaining required ingredients are the
relative numbers of thin disk, thick disk, and halo stars.

Given these assumptions, the equations establishing the probability
that a star belongs to the thin disk ($P_{thin}$), the thick disk
($P_{thick}$), or the halo ($P_{halo}$) are

\begin{equation}
P_{thin} = f_1{ {P_1\over{P}}},~~~~ P_{thick} =f_2{ {P_2\over{P}}}, ~~~~ P_{halo} =f_3{ {P_3\over{P}}} \\
\end{equation}

\noindent where,

\begin{eqnarray}
P & = & \sum f_i P_i, \nonumber \\
   & & 	\nonumber	   \\
P_{i}   & = & K_i \times {exp \left[{  -{{U^{2}_{LSR}}\over{2\sigma^{2}_{U_{i}}}}- {{(V_{LSR}-V_{ad})^{2}}\over{2\sigma^{2}_{V_{i}}}}
     - {{W^{2}_{LSR}}\over{2\sigma^{2}_{W_{i}}}}  } \right]   }  \nonumber \\
{\rm and} & & \nonumber  \\
K_i & = & {1\over{(2\pi)^{\frac{3}{2}} \sigma_{U_{i}} \sigma_{V_{i}} \sigma_{W_{i}}}} ~~~(i=1, 2, 3)\nonumber \\
\end{eqnarray}

\noindent and
$V_{ad}$ is the asymmetric drift, the mean galactic rotation velocity for
each stellar population relative to the LSR.

The parameters ($\sigma$'s and the mean velocities)
defining the Gaussian distribution functions  and the population
fractions $f_i$ given in Table~1  are taken from Robin et al.
(2003; but see also Ojha et al. 1996, 1999 and Soubiran et al. 2003).
The  asymmetric drifts $V_{ad}$ given by Robin et al. are refered to the Sun,
and therefore we have corrected it for the solar motion relative to the LSR
$V_{\odot}$ = +5.3 km s$^{-1}$.
The mean values of $U_{LSR}$ and $W_{LSR}$ for any of the three
populations are taken to be zero.
For the thin disk, the  estimates refer to stars of
5-7 Gyr in age, which is an average value for our sample (see \S \ref{ages}),
but this result is also in excellent agreement with the value inferred
by Allende Prieto et al. (2004) from FGK stars in the solar neighborhood.

The relative numbers of thin disk, thick disk, and halo stars
in the solar vicinity are taken to be
$f_{1} = 0.93$, $f_{2} = 0.07$, and $f_{3} = 0.006$, respectively.
As a check, we have  used
Nordstr\"{o}m et al.'s (2004) survey of Galactic F-G dwarfs in the solar
neighborhood
to estimate the fractions of the three components.
If we assume that all the stars having
$V_{LSR}$  between $-$150 and $-$40 km s$^{-1}$ belong to the thick disk,
 stars with $V_{LSR}$ $>$ $-$40 km s$^{-1}$
are part of the thin disk, and the rest are  halo stars, we find  fractions
of 94$\%$, 5$\%$, and 1$\%$.
The differences between
Robin et al.'s  (2003) and these fractions are small and
have no impact on the resolution of the sample into three components
based on probabilities.\footnote{Venn et al. (2004) seem to
have computed membership probabilities on the
assumption that the fractional populations are the same for the three components.
This assumption can give very different and misleading results.}
Our adopted fractions also compare well with the ratios
$ \rho_{thick}/\rho_{thin}$ of 4\% and 9\% derived at the solar radius
from star counts by Juric et al.
(2005) using SDSS and by Cabrera-Lavers et al. (2005) based on 2MASS,
respectively.

The  probabilities -- $P_{thin}, P_{thick},$ and $P_{halo}$ --
and the associated errors due to
errors in the velocities $U_{LSR}$, $V_{LSR}$ and $W_{LSR}$
were computed. (Our program reproduces the probabilities given by
Bensby et al. (2003, 2005) and Mishenina et al. (2004) when their
input data are adopted.)
We consider that a probability (P$_{thick}$ $-$ error) in excess
70$\%$ suffices to assign a star to
either the thin disk, the thick disk, or the
halo. 
Table~2 lists 95 thick stars with $P_{thick}$ given in column 12,
13 thin disk stars with $P_{thin}$ in column 12, and 20 halo stars
with $P_{halo}$ in column 12. Of the sample of 176 stars, there remain
48 stars with 34 belonging to either the thick or the thin disk with
about equal probability, and 14 belong to the thick disk or the halo with
approximately equal probability. In Table~2, $P_{thick}$ is given for
these 48 stars.
(The same procedure  applied to the 181 `thin' disk stars of Paper~I
yielded 175 thin disk,  just two thick stars  with the other four
assignable with roughly equal probability
to either the thin or the  thick disk.)

The present sample and that from Paper I are shown in Figure~1 with
the velocities $U_{LSR}, V_{LSR}$, and $W_{LSR}$ plotted against
[Fe/H] -- [Fe/H] for the present sample is given below. Also, shown is the
Toomre diagram with $(U_{LSR}^2 + W_{LSR}^2)^{1/2}$ against $V_{LSR}$.
The biases from
the  selection criteria for the present sample -- $V_{LSR} \leq -40$ km s$^{-1}$
and $\left|W_{LSR}\right| \geq 30$ km s$^{-1}$ -- are quite evident in the
three panels involving [Fe/H].  In addition, the sample from Paper~I was
biased towards less than solar metallicities, and, therefore, the thin disk
is underrepresented at [Fe/H] $\sim 0$.
The stars with approximately equal probability
of membership of the thin or thick disks fall, as expected, at the
thin-thick disk boundaries in the four panels.

Classification by probability with the parameters in Table~1 applied to our
sample selected by our $V_{LSR}$ and $W_{LSR}$ criteria results in
a collection of thin disk stars that may be deemed unusual by purists.
The computed orbits of the thin disk stars are quite eccentric;
compare the thin disk entries in Table~1 with those in Table~1 from
Paper~I -- also see Figure~1 or the different locations of these
thin disk stars and those from Paper~I.  The probability $P_{thin} > 70$\%
is largely set by the low $\left|W_{LSR}\right|$  which in turn implies
the star remains close to the Galactic plane. In several of the panels of
Figures~1 and~3, these thin disk stars fall at a boundary between
the thin disk stars of Paper~I and the present thick disk stars. Thus, one might
regard them on kinematic grounds as either extreme representatives
of the thin disk or the thick disk. Their compositions may resolve this
ambiguity (see below).

An informative illustration of the dependence of $P_{thick}$ on
the velocity components $U_{LSR}$, $V_{LSR}$, and $W_{LSR}$ is
provided in Figure~2.
This gives
equal probability contours for $P_{thick}$
equal to 50, 70, 80, and $\geq$90 per cent
for the range of $|U_{LSR}| =$ 0 to 200 km s$^{-1}$, $V_{LSR} =$ $+$100 to $-$200 km s$^{-1}$, and
$|W_{LSR}| = 0$ to 150 km s$^{-1}$.
In the figure, probabilities increase from outside ($P_{thick}$=0.50)
to inside ($P_{thick}$=0.97 for panels
(a) to (c) and to $P_{thick}$ = 0.90 for the panel (d)).
In these panels,  if a star's
$V_{LSR}$ and $W_{LSR}$ is such that it falls outside
of the thick disk $P_{thick}$=0.50 contour, the star  most likely
belongs to the halo and it is most likely a thin disk star if
it falls in empty bottom right corners of the plots.

\subsection{Orbital Parameters}
Orbital parameters such as the  eccentricity, maximum
distance above the Galactic plane, the apogalactic and the perigalactic
distance
for the sample were computed using the same Galactic potential model adopted for
Paper I.
The distance of 8.5 kpc for the Sun from the Galactic
center is adopted. The mean Galactocentric distance ($R_{m}$) for each star
is taken as the mean of the apogalactic and perigalactic distances.
Key Galactic orbital parameters
for the thick disk sample
are shown in Figure~3 against [Fe/H] for both the current sample and that from
Paper I. Again, it must be recognized that our selection criteria affect the
distribution of points in the panels of Figure~3.

\section{Observations}

High-resolution spectra of the programme stars were obtained  during
the period, December 2002 - June 2004 at
the Harlan J. Smith 2.7-m telescope of the W. J. McDonald Observatory,
using the 2dcoud\'{e} echelle spectrometer (Tull et al. 1995)
with a 2048 $\times$ 2048 Tektronix CCD as detector.

Spectral coverage at a resolving power of about 60,000 was complete
from  3500 \AA\ to 5600 \AA\ and substantial but incomplete from 5600 \AA\ to
about 9000 \AA. The Echelle spectroscopic data were reduced to one dimensional
spectra with Y-axis as normalized flux and X-axis as wavelength using
spectral reduction programme $IRAF$\footnote{IRAF is distributed by the National
Optical Astronomical Observatories, which is operated by the
Association for Universities for Research in Astronomy, Inc., under
contract to the National Science Foundation.}
as outlined in Paper~I. The final
reduced spectra have S/N$\approx$ 100 - 200.
Selection and measurement of suitable absorption lines followed the
procedures described in Paper I.

\section{Analysis}

The LTE abundance analysis was modelled as closely as possible
on that described in
Paper~I. The model atmosphere grid and the methods of determining
the fundamental atmospheric parameters were retained. A minor
alteration in the determination of the effective temperature
for about 40 stars is noted below. The line list and basic atomic
data was taken over from Paper~I.
The abundance analysis was again performed with the 2002 version of
the code MOOG (Sneden 1973).
The reader interested in the details is referred to Paper~I.

The distribution of stars in our sample with $T_{\rm eff}$, $\log$ g, and
[Fe/H]  is shown in Figure~4 where  is also given the distribution
of $\log$ g with [Fe/H]. This figure may be compared with its
counterpart in Paper~I. The comparison shows the anticipated difference in
the [Fe/H] range of the two samples. A point of note is that the
samples differ in the spans of $T_{\rm eff}$; the thin disk stars
are in the mean systematically warmer than the present sample
with the peak in the thick disk distribution being about 300~K cooler
than that in Paper I.

\subsection{The Effective Temperature}

In Paper~I, effective temperatures ($T_{\rm eff}$) were estimated from
a star's Str\"{o}mgren $(b-y)$ colour using the calibration provided
by Alonso et al. (1996) who obtained $T_{\rm eff}$ from the infra-red
flux method (IRFM). The $uvby\beta$ data was adopted from
Hauck $\&$ Mermilliod's (1997) catalogue.
This approach was used for  the 135 of the 176 stars in the
present sample  for which Str\"{o}mgren photometry has been
reported.

An alternative approach was developed for the other 41 stars.
We chose to use the $(V-K)$ colour and the corresponding
calibration, again from Alonso et al.
The $V$ magnitude was taken from  the {\it Hipparcos} catalogue.
The  $K$ magnitude is not available for all of these
stars, but the
2MASS Catalogue\footnote{This publication makes use of data products from the Two Micron All
    Sky Survey, which is a joint project of the University of Massachusetts
    and the Infrared Processing and Analysis Center/California Institute of
    Technology, funded by the National Aeronautics and Space Administration
    and the National Science Foundation} (Cutri et al. 2003)  provides the
magnitudes $K_s$ for all 41 stars.
To estimate a possible
systematic offset between the $(K_s)$ magnitudes of 2MASS and the
Alonso et al. scale ($K$), we took a
sample of 100 stars from the 2MASS catalogue
with observations from Alonso et al (1996). The mean difference
is very small: $(K_s-K) = -0.004$ with a $\sigma$ of 0.09 and no
trend (Figure~5). Therefore, we adopted the 2MASS $K_s$ and equated it with
$K$.

The $(b-y)$ and $(V-K) \equiv (V-K_s)$ colours give very similar
$T_{\rm eff}$ values -- see Figure~6. The mean difference
$T_{\rm eff} (b-y) - T_{\rm eff}(V-K)$ = $-$15~K with a $\sigma = 69$~K
with a few outliers. For the outliers, we obtained $T_{\rm eff}$
spectroscopically by demanding excitation equilibrium for a set of
Fe\,{\sc i} lines.

Observed colours were not corrected for interstellar extinction
for stars closer than 100 pc. For the 37 stars with distances
greater than 100 pc (but less than 150 pc), a correction was
estimated from Neckel, Klare \& Sarcander (1980) maps. The maximum
extinction is about 0.1 magnitudes in $V$ and thus 0.01 magnitudes in
$K$. Correction for this level of extinction, a maximum for our
sample, increases the $T_{\rm eff}$ by about 140 K.

\subsection {Metallicity, surface gravities, and microturbulence}

Metallicity  usually refers to the iron abundance
(relative to the solar abundance) which here was obtained for each star
from numerous well defined
Fe\,{\sc i} lines and  a few Fe\,{\sc ii} lines.
Since the stellar $T_{\rm eff}$
 except for a few exceptional cases (see above) and the surface gravity in all
cases were obtained without recourse to the Fe\,{\sc i} and Fe\,{\sc ii}
lines, it is of interest to see if the Fe\,{\sc i} and Fe\,{\sc ii}
lines return the same value for the iron abundance. Non-LTE effects
may affect these iron abundance estimates; overionization of neutral iron
(relative to LTE) is believed to be the principal non-LTE effect
so that, in an LTE analysis, the abundance from Fe\,{\sc i} lines is
less than that from Fe\,{\sc ii} lines. Figure~7 (top panel)
shows the  abundance difference
from Fe\,{\sc i} and Fe\,{\sc ii} as a function of the abundance from the
neutral lines. On average the Fe\,{\sc i} lines give a lower abundance
by 0.04$\pm$0.08 dex with no significant trend with [Fe/H]. (The
corresponding difference for the thin disk stars in Paper~I was
0.02$\pm$0.05 dex for stars with [Fe/H] from about 0.0 to $-$0.6.)
 Following
Paper~I, we adopt the iron abundance given by the Fe\,{\sc i} lines because
the neutral lines are many and the ionized lines few.

Photometric recipes exist for determining the metallicity, here identified
with [M/H]. Here, we adopt the metallicity calibration of Str\"{o}mgren
photometry provided by Hauck $\&$ Mermilliod (1997). Figure~7 (lower panel) shows the
difference between the photometric [M/H] and the spectroscopic [Fe/H]
from the Fe\,{\sc i} lines as a function of the spectroscopic [Fe/H].
The differences are generally small and the mean difference
[M/H] $-$ [Fe/H] $= -0.01\pm0.10$. (In Paper~I, the mean difference was
0.05$\pm$0.09 with no detectable trend over the interval
[Fe/H] from $-$0.2 to $-$0.8.)
We note that discrepant iron abundances are inferred
from Fe\,{\sc i} and Fe\,{\sc i} lines for slightly lower values of
$T_{\rm eff}$ and especially more metal rich stars than in our sample
(Feltzing \& Gustafsson 1998; Schuter et al. 2003; Yong et al.
2004; Allende Prieto et al. 2004).

Surface gravities were obtained, as  in Paper~I,
by a combination of stellar isochrones (Bertelli et al. 1994),
$T_{\rm eff}$, [Fe/H],
and the $Hipparcos$ astrometry. The resulting log $g$ values are compared for the
common stars between this study and few others (see Table~3).
For the
microturbulent velocities ($\xi_{t}$) we used the relation
between $\xi_{t}$, $T_{\rm eff}$, and log $g$ derived in Paper~I.
The relation used from Paper~I is derived using mainly thin disk stars
which cover slightly different ranges in the parameters than the thick disk sample
stars. To check the validity of the relation we derived $\xi_{t}$ values
for 15 thick disk stars using Fe~I lines. The adopted values from the relation
are larger by only 0.14$\pm$0.17 km s$^{-1}$ than the spectroscopically derived values.
The effect of such difference on the abundances is negligible.

\subsection{Model Atmospheres}

Stellar abundances  are obtained assuming LTE line formation. Abundances
are given with respect to the Sun which was analyzed as described in
Paper~I.
We used Kurucz's (1998)   LTE, plane parallel, line-blanketed models
with convective overshoot and
the revised (2002)
stellar abundance code MOOG (Sneden 1973). The oscillator
 strengths for the basic set of about
160 lines are a mixture of laboratory measured values and the astrophysically
derived by inverting solar and the stellar spectra (see Paper~I).

The rationale for adopting Kurucz  models with convective overshoot
is described in Paper I; the thin disk sample in Paper~I are similar to the
Sun for which the convective overshoot model gives a good
representation.
However, the current thick
disk
stars are cooler, metal-poor and higher
gravity stars than the thin disk sample; they are not close
analogs of the Sun.
The models with convective overshoot are not widely used
in abundance analysis.
For this reason, we computed the  [X/Fe] of 12 representative elements
using
both models with and without convective overshoot.
Twelve stars are
chosen so that
they cover $T_{\rm eff}$ (5000~K to 6100~K) and [Fe/H] (0.2 to $-$1.6)
range of the thick disk sample.
Differences  ($\delta$[X/Fe]) in [X/Fe] from the two kinds of models
were $\delta$[X/Fe] $\leq$ 0.01 for all the
elements except
carbon for which 0.04 was representative.

\subsection{Abundance errors}

The principal goal of this study is to define
the differences in composition between thin
and thick disk stars. It is known that the differences,
even in the most striking cases, are small ($\leq 0.2$ dex) and, therefore,
attention must be paid to the errors thought to affect the
abundance determinations. In addition, our sample of thick disk stars
is of a sufficient size that we may attempt to estimate the
dispersion in the abundances at a given metallicity and wonder if
the dispersion is due to measurement errors or may contain
cosmic scatter.

The recipe used to assess the errors follows that used in Paper I.
Two qualitative differences may be noted. First, the
spectra for the thick disk sample  are of lower quality than those
used for Paper I: $S/N \simeq 100$ -- $200$ versus $S/N \simeq
$200 -- 400. Also, two spectra were generally obtained for each thin
disk star but a single one for a thick disk star; the thick disk stars
are relatively fainter than the selected thin disk stars.
Second,  the thick disk stars
are systematically cooler (and more metal poor) than the thin
disk stars; these differences lead to different sensitivities of the
abundances on the atmospheric parameters.

For these reasons, we estimated uncertainties in [X/Fe] due to uncertainties in
model parameters and measurement errors afresh for the present sample.
We assumed that the errors are uncorrelated.
The predicted uncertainty, $\sigma_{mod}$ for a whole sample,  in each abundance
ratio due to measurement errors can be written as a simple mean of the $\sigma$s
estimated for $n$ representative stars

\begin{equation}
\sigma_{\rm mod}  =   \frac{1}{n} \sum_i \sigma_i
= \frac{1}{n} \sum_i
\sqrt{\sum_j \left(\frac{\partial}{\partial p_j} {\rm [X/Fe]}\right)^2
\left(\delta p_j\right)^2
} \nonumber \\
\end{equation}

\noindent and the parameters $p_j$ considered here
are the effective temperature,
surface gravity, metallicity, microturbulence, and the equivalent width
measurements.
We examined five representative thick disk stars spanning the
sample's  range in $T_{\rm eff}$, [Fe/H],  and $\log$ g.
Assuming that  $\delta T_{\rm eff}$ = 100~K,
$\delta$log $g$ = 0.2, $\delta$[M/H] = 0.2, $\delta$$\xi_{t}$ = 0.25 km s$^{-1}$,
and $\delta W_{\lambda}$ = 2~m\AA.
The error in $W_{\lambda}$ is estimated in the
same way as in Paper~I, and the resulting error
in the abundance is divided by $\sqrt{N}$ where $N$ is the number of lines
used in deriving the abundance [X/H].
In Table~4, $\sigma$s for each abundance ratio for five
stars and the final mean value $\sigma_{mod}$ and the
standard deviation are given.

\renewcommand{\thetable}{4}
\begin{table*}
\centering
\caption{Abundance uncertainties due to estimated uncertainties
in atmospheric parameters for five representative stars.
The $\sigma$'s are quadratic sum of variations in
abundance ratios, [X/Fe], due to uncertainties in model parameters. The column $\sigma_{mod}$,
is the mean of the $\sigma$'s and the quoted error $std$ is the standard deviation per measurement. }
\begin{tabular}{@{}lcccccc@{}}
\hline \hline
            & HIP\,17666  & HIP\,15405  & HIP\,5122 & HIP\,3086 & HIP\,16738
 &   \\
$T_{\rm eff}$&  5000~K & 5107~K     & 5204~K   & 5697~K   & 6000~K                      \\
\lbrack Fe/H\rbrack & $-$1.03 & $-$0.73 & $-$0.55 & $-$0.23 & $+$0.36   &  \\
& $\sigma_{1}$ & $\sigma_{2}$& $\sigma_{3}$&$\sigma_{4}$& $\sigma_{5}$& $\sigma_{mod}\pm{std}$  \\
\hline
\lbrack Fe/H\rbrack  &0.07  & 0.09  &0.09  & 0.08 & 0.09  & 0.08$\pm$0.01   \\
\lbrack C/Fe\rbrack  &0.07  & 0.21 &0.10  & 0.16 &0.16  & 0.14$\pm$0.06   \\
\lbrack O/Fe\rbrack  &0.21  & 0.21 &0.20  & ... & 0.15 & 0.19$\pm$0.03   \\
\lbrack Na/Fe\rbrack &0.06  & 0.05 &0.04  & 0.03  &0.04  & 0.05$\pm$0.02   \\
\lbrack Mg/Fe\rbrack &0.05  & 0.05 &0.04  & 0.04 &0.06  & 0.05$\pm$0.01   \\
\lbrack Al/Fe\rbrack &0.04  & 0.05 &0.05  & 0.05 &0.05  & 0.05$\pm$0.01   \\
\lbrack Si/Fe\rbrack&0.09  & 0.10 &0.08  & 0.03 & 0.06 & 0.07$\pm$0.03   \\
\lbrack S/Fe\rbrack  &0.07  &0.18  &0.17  & 0.12  &0.13  & 0.13$\pm$0.04    \\
\lbrack Ca/Fe\rbrack &0.07  &0.04  &0.03  & 0.02 &0.01  & 0.03$\pm$0.03   \\
\lbrack Sc/Fe\rbrack &0.10  &0.13  &0.11  & 0.09 &0.13  & 0.11$\pm$0.02   \\
\lbrack Ti/Fe\rbrack &0.09  &0.07  &0.06  & 0.04 &0.04  & 0.06$\pm$0.02  \\
\lbrack V/Fe\rbrack  &0.10  &0.07  &0.05  & 0.03  &0.02  & 0.06$\pm$0.03   \\
\lbrack Cr/Fe\rbrack &0.05  &0.03  &0.03  & 0.02  &0.02  & 0.03$\pm$0.01   \\
\lbrack Mn/Fe\rbrack &0.05  &0.08  &0.07  & 0.05 & 0.04 & 0.06$\pm$0.02   \\
\lbrack Co/Fe\rbrack &0.02  &0.03  &0.03  & 0.03  &0.02  & 0.03$\pm$0.01   \\
\lbrack Ni/Fe\rbrack &0.05  &0.04  &0.03  & 0.03 & 0.04 & 0.04$\pm$0.01   \\
\lbrack Cu/Fe\rbrack &0.05  &0.03  &0.04  & 0.05 & 0.03 & 0.04$\pm$0.01   \\
\lbrack Zn/Fe\rbrack &0.09  &0.11  &0.11  & 0.07 & 0.10 & 0.10$\pm$0.02   \\
\lbrack Y/Fe\rbrack  &0.09  &0.12  &0.13  & 0.12 &0.07  & 0.11$\pm$0.03   \\
\lbrack Ba/Fe\rbrack &0.11  &0.11  &0.13  & 0.13 &0.13  & 0.12$\pm$0.01   \\
\lbrack Ce/Fe\rbrack &0.10  &0.11  &0.12  & 0.11  &...  & 0.11$\pm$0.01   \\
\lbrack Nd/Fe\rbrack &0.08  &0.11  &0.13  & 0.12  &...  & 0.11$\pm$0.02   \\
\lbrack Eu/Fe\rbrack &0.12  &...  &0.14  & 0.13  &...  & 0.13$\pm$0.01   \\
\hline
\end{tabular}
\end{table*}

\subsection{Ages}
\label{ages}

Stellar ages for the present sample have been computed by the method used
in Paper~I. Briefly, we used the stellar isochrones of Bertelli et al. (1994)
and the method described in Paper~I and Allende Prieto et al. (2004).
The adopted isochrones do not consider compositions in which the
$\alpha$-elements have enhanced abundances (relative to iron). For
thin disk stars, this is a fair approximation.
However, in the present sample there are stars belonging to the thick
disk and in halo which show $\alpha$/Fe ratios significantly different
from zero. Neglecting the $\alpha$-enhancements  in the isochrones
would overestimate the ages for the most metal-poor thick disk and
halo stars by up to 2 Gyr.
We thus adopted the relationship between metallicity ($Z$), iron abundance
([Fe/H]), and the $\alpha$-enhancement proposed by Degl'Innocenti,
Prada Moroni \& Ricci (2005), in selecting the appropriate isochrones
from the Padova grid.
Reliable ages are determinable only for those stars which have evolved away
from the zero age main sequence. Ages were estimated for 65 stars
(see Table~2, column 11).

In Figure~8, we compare our age determinations
with those published in the recent survey by Nordstr\"{o}m et al. (2004).
Nordstr\"{o}m et al.
used the Padova isochrones (no $\alpha$-enhancement) of Girardi et al. (2000).
The 45 stars which are in common between the two surveys are overwhelmingly
thin disk stars, and therefore,
in spite of the difference in the selection of the isochrones' metallicity,
the two determinations agree fairly well with
an average difference of 0.1 Gyrs with $\sigma \approx 2$ Gyrs.

\section{Thin and thick disk compositions}

\subsection{Some Comparisons}

Clear evidence that thick and thin disk stars of the
same [Fe/H] showed different abundances (relative to Fe) of
other elements was provided by Furhrmann's (1998) demonstration that
[Mg/Fe] was systematically greater in the thick disk stars.
Extension of this result to other elements was  made
by Prochaska et al. (2000), who determined the abundances of up to
20 elements in a sample  of 10 stars with
$V_{LSR}$ from $-20$ to $-100$ km s$^{-1}$ and a $W_{LSR}$ that
takes a star to at least 600 pc above the Galactic plane, which
virtually guarantees that the stars belong to the thick disk.

Prochaska et al.'s survey augmented by the occasional inclusion
of thick disk stars in quite extensive studies of local stars
(Edvardsson et al. 1993;  Chen et al. 2000; Fulbright 2000;
Reddy et al. 2003)
led to the finding that abundance differences between thick and thin
disk stars appear to define two broad categories.
In the first category (here, Mg-like elements)
are elements like Mg in which [X/Fe] for a thick
disk star exceeds that in a thin disk star of the same [Fe/H].
The thick-thin difference is not the same for all elements
in this category and may also be a function of [Fe/H].
The second category (here, Ni-like elements)
are those elements for which [X/Fe] appears unchanged
between the thin and thick disk. In reality, the thick-thin disk
abundance differences may span a continuous range.

Before presenting and discussing our results in detail, we
offer a few comparisons with several recent abundance
analyses of thick disk stars to highlight differences in
the sizes of the samples and to examine the consistency
between the  different investigations with respect to
the abundance ratios. The chosen analyses are those by
Bensby and colleagues, Mishenina et al. (2004), and
Fuhrmann (2004). For each case, we reassessed the
assignments to the thick and thin disks using our
recipe for the membership probabilities.

Bensby  and
colleagues determined abundances for many elements
in  a sample of 102 F-G dwarfs of which 35 were
attributed by them to
the thick disk. According to our
recipe,  18 of the 102 are thick disk stars.
Though the recipes for choosing thick disk stars are
basically the same, they differ in the normalization.
Bensby et al. chose a relative probability ratio
of thick disk and thin disk  $>$
1.0, and halo and thick disk ratio $<$ 1.0 for a star to belong thick disk.
Their criteria
approximately translate into our criteria if $P_{thick}$ $>$ 50$\%$
for a star to belong thick disk.

Mishenina et al. analysed 174 F-G-K dwarfs for their
Mg, Si, Fe, and Ni abundances. (Non-LTE effects
were considered in the Mg analysis.) Thirty stars were
assigned to the thick disk. Adoption of our method of
calculating the membership probability reduces the number of thick disk
stars to 13. The main difference between us and Mishenina et al.
is in the adopted fraction for thin and thick disk stars. Mishenina et al. assumed
25$\%$ thick disk and 75$\%$ thin disk stars. This led to
larger number of thick disk stars in their study.

\renewcommand{\thetable}{3}
\begin{table}
\centering
\caption{ Mean differences and standard deviations of the
abundance ratios [X/Fe] for stars that are common among
Bensby et al. (2003,2004), and Mishenina et al. (2004).}

\begin{tabular}{@{}lrrrr@{}}
\hline \hline
         &\multicolumn{2}{c}{Bensby - Ours} &\multicolumn{2}{c}{Mishenina- Ours}\\
Quantity & diff.  & $\sigma$ &  diff.    & $\sigma$ \\
\hline
T$_{\rm eff}$& 88  &  80    &   102      & 116   \\
log $g$     & $-$0.05 &  0.08    &   $-$0.18      & 0.18 \\
${\rm [Fe/H]}$   & 0.03     & 0.05     &   $-$0.03        &0.09    \\
${\rm[Na/Fe]}$  & -0.02  & 0.05     &   0.03        &  0.08        \\
${\rm[Mg/Fe]}$  & 0.00     & 0.06     & 0.05          & 0.05        \\
${\rm[Al/Fe]}$  & 0.04     & 0.08     & 0.12          & 0.10       \\
${\rm[Si/Fe]}$  & 0.00     & 0.06     & 0.16          & 0.09       \\
${\rm[Ca/Fe]}$  & 0.04     & 0.06     & 0.07          & 0.04        \\
${\rm[Ti/Fe]}$  & 0.08     & 0.04     & 0.10          & 0.08       \\
${\rm[V/Fe]}$   & ...      & ...      & 0.05          & 0.09       \\
${\rm[Cr/Fe]}$  & 0.04     & 0.05     & 0.08          & 0.03       \\
${\rm[Ni/Fe]}$  & 0.01     & 0.04     & 0.05          & 0.02       \\
${\rm[Zn/Fe]}$  & 0.02     & 0.05     & ...           & ...        \\
${\rm[Y/Fe]}$   & -0.08    & 0.04     & 0.01          & 0.05       \\
${\rm[Ba/Fe]}$  & 0.12     & 0.08     & 0.12          &  0.10       \\
\hline

\end{tabular}
\end{table}

Fuhrmann (2004) following his earlier work (Fuhrmann 1998)
obtained Mg and Fe abundances for a sample of 71 nearby stars. Out of which
25 or more are deemed thick disk stars.

In Figure~9, we show  [Mg/Fe] against [Fe/H] as provided by our
sample and the three published analyses. The larger filled black symbols
in the plot represent thick disk stars
based on our criteria.
Stars for which the probability of thin or
thick disk membership is below 70\% but greater that 50$\%$ are included in the panels
but with a smaller symbol.
Figures~10 and 11 show the
corresponding results for [Si/Fe], and [Ti/Fe], respectively.
These figures show clearly the increase in the number of thick disk stars
provided by our sample. They also show
broad agreement between the different studies over the behaviour of Mg-like
elements with [Fe/H] in thin and thick disks: [X/Fe] of the thick disk
exceeds that of the thin disk at a common [Fe/H] for [Fe/H] $\leq -0.3$
with an apparent merger of thin and thick disk behaviour for greater
[Fe/H]. Inspection of Figures~9, 10, and 11 shows that
a majority of stars from Bensby et al'.s sample
which we put in the thin-thick group with [Fe/H] $\leq$ $-$0.3
are thick disk stars as judged by Mg, Si, and Ti abundances.
Mishenina et al.'s results give marginal support to the
difference in [Mg/Fe]  and [Si/Fe] between thin and thick disks.

A comparison of abundances for stars in common shows that the different
studies are consistent.
We have ten stars (no distinction is made according to thin or thick
disk)
in common with Bensby and colleagues' total sample of 102 stars.
We have five stars in
common with Mishenina et al. (2004).
Mean difference and standard deviations for the samples
of common stars are summarised in Table~3 for the
atmospheric parameters, the Fe abundance, and the [X/Fe]
values.
We have six stars in common with Fuhrmann (2004). The mean difference
between him and us, for [Fe/H] is 0.0$\pm$0.05 and [Mg/Fe] is 0.03$\pm$0.12.
Fuhrmann's $T_{\rm eff}$ and log $g$ values are systematically
hotter by 125$\pm$23~K and lower by 0.19$\pm$0.06~dex, respectively.
Brewer \& Carney (2005) analyzed 23 G dwarfs of which four were
analyzed by us. Their and our adopted atmospheric parameters and
derived abundances are in excellent agreement, e.g. differences
in $T_{\rm eff}$, $\log$ g, and [Fe/H] run from $-43$ to $+14$ K, $-0.2$ to
0.0 cgs units, and $-0.08$ to $+0.06$ dex, respectively.

An analogous and equally satisfactory
comparison was made in Paper I for thin disk stars and
the large surveys by Edvardsson et al. (1993), Chen et al.
(2000), and Fulbright (2000). These analyses of stars in common with our
studies suggest that results from different studies could
be combined to enlarge the sample size for thick and thin disk
stars.

\subsection {Thin and Thin-Thick Disk stars}

Our discussion of the thick disk's composition is based on the
95 thick stars chosen by the requirement that $P_{thick} > 70$\%.
Here, we consider the
13 stars assigned to the thin
disk and the 34 stars in the thin-thick disk category.
Figure~12 shows in separate panels the run of [$\alpha$/Fe] ratio
versus [Fe/H] for thick, thin-thick, and thin disk stars
from our present sample with the thin disk stars from
Paper I shown in each panel.
The ratio [$\alpha$/Fe] is defined as simple mean of
the four ratios [Mg/Fe], [Si/Fe], [Ca/Fe], and [Ti/Fe].
Clearly, the samples of thick/thin, and thin disk stars both
contain thick stars, as judged by [$\alpha$/Fe].
This mixing and contamination
must in part result from the fact that the assignment to
the thin or the thick disk is based on a probability. Given that
Figure~12 (and other figures) convey the impression that thin and
thick disk abundance relations are distinct,
the collection
of thick disk stars might be augmented by stars from the thin-thick
disk category, and even the thin disk category. We do not act on the
second suggestion, but reconsideration of the procedure for
categorising disk stars would be a useful exercise.

\subsection{Our halo stars}

The probability criteria
yielded 20 halo stars. Most of them, as expected, have very negative
$V_{LSR}$ $\leq$ $-$180 km s$^{-1}$
and orbits of  extreme eccentricity  (0.7 to 1.0).
A few stars are on retrograde orbits,
i.e., $V_{LSR}$ $<$ $-$220 km s$^{-1}$.
The halo stars in the sample
span the [Fe/H] range $-$0.6 to $-$2.0.

Mg-like elements show overabundances
similar to thick disk stars and may show a mild
trend of [X/Fe]  with [Fe/H]. At overlapping metallicities, the
[$\alpha$/Fe] of the halo stars are in agreement with the
thick disk ratios. Ni-like elements for the halo stars track the Fe
similar to thin and thick disk
stars. The scatter in [X/Fe] ratios for both the Mg-like and
Ni-like elements for the halo stars appears similar to that for
 the thick disk stars.

The [$\alpha$/Fe] ratio,
for the halo stars in our sample is about
0.25 dex, in agreement with our thick disk stars.
This value is in good agreement with the [$\alpha$/Fe] results for halo stars
from Gratton et al. (2003) in the same [Fe/H] range. Their
results also do not show a trend of [$\alpha$/Fe] with [Fe/H].

\subsection{The Mg-like elements}

One subset of the Mg-like elements are the so-called $\alpha$-elements:
C, O, Mg, Si, Ca, and Ti.
Our results for these elements are shown in
Figures~13 and 14.
Other Mg-like elements are Al, Sc, V, Co, and possibly Zn (Figure~15).

The trend of [X/Fe] against [Fe/H] for Mg-like elements in
 the thick disk may be
characterized by a shallow slope for [Fe/H] $\leq -0.3$  and
 from the trend for thin disk stars.
For each Mg-like element, we compute the slope
($B_{thick}$) from a linear regression fit to the [X/Fe] from
[Fe/H] $ = -0.3$ to $-1.0$. The offset between the thick and thin trends
is obtained by calculating the mean value ($M_{thick}$ or $M_{thin}$)
in the bin of [Fe/H] from $-0.45$ to $-$0.55.
Quantities $B_{thin}$, $B_{thick}$, $M_{thin}$
and $M_{thick}$ are given in Table~6 together with the
thick-thin disk offset $\Delta[X/Fe] = M_{thick} - M_{thin}$.
The behavior of the Mg-like elements for [Fe/H] $> -0.3$ is
discussed below.
Below we make  few remarks on individual elements.

{\bf Carbon:} We obtained abundances for stars with [Fe/H] $\geq$ $-$1.2;
the chosen
C\,{\sc i} lines are too weak to measure in more metal-poor stars.
Abundance ratios [C/Fe] for the thick disk sample
stars below [Fe/H] $< -$0.4 as shown
in Figure~13 are on average larger than the thin disk stars of the
same  [Fe/H]. Although the scatter is quite large, carbon is seen to
behave like Mg and other $\alpha$-elements.
Similar behaviour of [C/Fe] with [Fe/H] is seen by
Shi, Zhao \& Chen (2002) and Takeda-Hidai et al. (2005)
for nearby F, G, and K dwarfs. They also noted
that the non-LTE corrections are negligibly small.
Our thin disk [C/Fe] (Paper~I), also from the C\,{\sc i} lines, are in
 agreement with results
obtained from [C\,{\sc i}] line at 8727~\AA\  by
Gustafsson et al. (1999) and Allende Prieto et al. (2004)
for samples dominated by thin disk stars.
Contrary to the above studies, the recent study by Bensby et al. (2005)
based on the [C\,{\sc i}] line at 8727~\AA\
 for a small sample of both thick and
thin disk stars
shows almost a flat trend of [C/Fe] below [Fe/H] $\approx$ $-$0.2.

{\bf Oxygen:}
The permitted O\,{\sc i}
 lines at 7771~\AA\ are used to get O abundances for all stars.
These lines are known to be subject to non-LTE effects.
We apply the empirical calibration to correct for these effects that was
developed in Paper~I using abundances from the [O\,{\sc i}] and O\,{\sc i}
lines for the same stars.
The resulting  [O/Fe] ratios are shown against [Fe/H], which
shows oxygen to behave as a Mg-like element (Figure~13).
Full non-LTE corrected O abundances will be discussed in a separate
publication (see
Ram\'{\i}rez, Allende Prieto, $\&$ Lambert, 2005).

{\bf The $\alpha$-elements: Mg, Si, Ca, and Ti:}
The runs of abundance ratios [X/Fe] against [Fe/H] are shown in Figure~14.
For [Fe/H] $< -$0.3, [X/Fe] for
thick disk stars are distinctly larger than for thin disk stars.
In the [Fe/H] range $-$0.3 to $-$0.6 where thin and thick disk stars
are well represente, there is a clear separation between [$\alpha$/Fe] of
Mg, Si, Ti, and probably also Ca for thin and thick disk stars.
For metal-rich stars of [Fe/H] $\geq$ $-$0.3, the thin and thick disk stars
show similar abundances.
We comment below (see Sec 6)
on the possibility that there is a smooth transition for [X/Fe] from
a positive displacement above its thin disk value for [Fe/H] $< -0.3$
to the thin disk value for [Fe/H] $> 0.0$.
 The thick disk [X/Fe] ratios smoothly merge with the halo ratios
in the overlapping metallicity range.

Our results for the $\alpha$-elements in the thick disk agree very well
with
Prochaska et al.'s (2000) results for their sample of ten stars with
[Fe/H] between $-$0.5 and  $-$1.2.
 Our results  also agree with
the results from Fuhrmann (1998, 2004) for Mg and
Bensby et al. (2005) for Mg, Si, Ca, and Ti (see Figures 9, 10, and 11).

{\bf Aluminium:}
An odd-Z  element Al might have been
expected to behave like Na, another odd-Z light element.
This expectation is not met (Figure~15); Al is Mg-like and Na is Ni-like.
Our results below solar metallicity agree with those by Prochaska
et al. (2000) and
Bensby et al. (2005).
Above [Fe/H] = 0.0, Bensby et al., also Chen et al. (2000),
 show  [Al/Fe]
increasing with increasing [Fe/H]. Our samples contain too few metal-rich stars to
detect this interesting trend.
Below the metal-poor end of our sample,  [Al/Fe] ratios for halo stars
are seen to
be sharply decreasing with decreasing [Fe/H]
(e.g; Spite \& Spite 1980; Gehren et al. 2004).

{\bf Scandium:}
The ratio [Sc/Fe] with [Fe/H] shows an interesting trend.
Above [Fe/H] = $-$0.3, [Sc/Fe] ratios are similar
for  thin and thick disk stars.
Below [Fe/H] = $-$0.3, [Sc/Fe] appears constant with an excess - albeit slight -
above thin disk values, a characteristic of a Mg-like element. A
 keen eye
may suggest that [Sc/Fe] slowly increases and then decreases at around
[Fe/H] $\approx$ $-$0.5 and returns to the solar ratio.

A similar trend of increasing [Sc/Fe] with [Fe/H] and then decreasing was
seen for
disk F- and G- dwarfs (Nissen et al. 2000; Prochaska and McWillam 2000).
For their thick disk stars with [Fe/H] range of $-$0.4 to $-$1.2,
Prochaska et al'.s  results agree with our larger data set
in magnitude and trend.

{\bf{Vanadium:}}
Below [Fe/H] = $-$0.3, [V/Fe]
of thick disk stars is slightly larger than for
thin disk stars.
In the [Fe/H] range $-$0.3 to $-$1.6,
[V/Fe] is almost flat with [Fe/H].
Our thick disk results are consistent with those from Prochaska et al. (2000).

Peterson (1981) derived [V/Fe] ratios for 22 moderately metal-poor
([Fe/H] = $-$0.5 to $-$1.5) F-G-K dwarfs. Her results show a
 clear and approximately uniform enhancement
of [V/Fe] relative to solar except for
two stars at the low [Fe/H] limit of the sample.
A quick look at the kinematics of her sample showed many of them
to have  $V_{LSR}$ $\leq$ $-$40 km s$^{-1}$,
 indicating membership
of the  thick disk or the  halo.

{\bf Cobalt:}
Figure~15 shows that cobalt is a Mg-like element.
For thick disk stars, [Co/Fe] is almost
flat with [Fe/H], in good agreement with the results obtained
by Prochaska et al. (2000). For thin-disk stars, a weak trend
of decreasing [Co/Fe] with increasing [Fe/H] is
apparent in the range $-0.5<$[Fe/H] $<0.0$. This is consistent
with the results of Allende Prieto et al. (2004) for that
interval (with the exception of a small scale offset).

{\bf Zinc:}
Zinc appears to be
indistinguishable between thin disk and thick disk stars in the
common metallicity range. We note, however, that 
for thin disk
stars there is evidence for a mild trend of increasing [Zn/Fe]
with decreasing [Fe/H], as previously reported by Nissen et al. (2004),
Allende Prieto et al. (2004), and Bensby et al. (2005).
The latter two studies also reported an increase of the Zn/Fe ratios at
[Fe/H] $>0$.

\subsection{The Ni-like elements}

The Ni-like elements include Na, Mn, Cr, Ni and Cu. Prochaska et al.
were the first to show that these elements (relative to Fe)
 behaved similarly in thick and
thin disk stars.
Bensby et al. (2005)  confirmed the Ni-like behavior of Na,
Cr and Ni but did not
consider either Mn or Cu in their analyses.
Our results for the quintet are shown in Figure~16.
Of especial
interest, perhaps, is the example of Mn where [Mn/Fe] is a quite
steeply increasing function of Fe, yet thin and thick disk stars
form a single relation. The contrast between the odd-even
Na (Ni-like) and Al (Mg-like) is also interesting.

{ \bf {Sodium:}}
This odd-Z and proton-capture element seems to be a difficult
one for which to obtain consistent values. We suspect departures
from LTE are behind these difficulties.
There are several studies but no two
are in very good agreement. For example, Chen et al.'s (2000)
study of disk F- and G- dwarfs shows a flat trend of
[Na/Fe] with [Fe/H] without Na enhancement. Edvardsson et al (1993) results
suggest [Na/Fe] increases with [Fe/H] for stars of [Fe/H] $>$ 0.0 and an apparent slow rise
in [Na/Fe] with decreasing [Fe/H]
for stars with [Fe/H] $<$ 0.0.
Prochaska et al. (2000) found an enhancement of Na for 10 thick disk stars without
any trend with [Fe/H].

The results from our study are based on two Na~I lines 6154~\AA\ and 6160~\AA.
Results indicate no visible (Figure~16)
demarcation between the thin and thick disk samples.
Examination of the plot indicates
a possible slow increase of [Na/Fe] with decreasing [Fe/H] reaching a maximum value of
[Na/Fe] $\approx$ 0.15 at [Fe/H] $\approx$ $-$0.6. From the maximum,
[Na/Fe] may fall with decreasing [Fe/H]. Recent analysis of disk dwarfs by Shi, Gehren, $\&$ Zhao (2004)
using the same lines as we did but with NLTE treament suggests a similar
trend of [Na/Fe] in [Fe/H] range 0.0 to $-$1.0.
For metal-rich stars, they confirm Edvardsson et al. (1993) results.

{\bf{Chromium and Nickel:}}
Chromium shows [X/Fe] $\simeq 0.02\pm$0.04 between thick and thin disk
stars.
Chromium tracks Fe well up to around [Fe/H] $\simeq -$0.6, but [Cr/Fe] is negative
for lower [Fe/H].
Nickel tracks Fe throughout the [Fe/H] range. Possibly,
 [Ni/Fe] is larger
for thick disk stars compared to thin disk but the mean difference
is just 0.04$\pm$0.03.
A small overabundance of Ni for the thick
disk stars is suggested by the equivalent figures given by
Prochaska et al. (2000) and Bensby et al. (2005).

{\bf {Manganese:}}
As shown in Figure~16, [Mn/Fe] is a function of [Fe/H]. [Mn/Fe] ratios
at a given [Fe/H] for thin, thick, and halo stars are the same.
 Below [Fe/H] = $-$1.0,
[Mn/Fe] shows a flat trend with [Fe/H]. Above [Fe/H] = $-$1.0, Mn steadily
increases with increasing [Fe/H].
Our results are in agreement with Mn results for disk and
halo stars (Nissen et al. 2000)
and the 10 thick disk stars analysed by Prochaska et al. (2000).

{\bf{Copper:}}
Copper is  a Ni-like element in the [Fe/H] interval populated by
both thin and thick disk stars; there is a hint that [Cu/Fe] is slightly
greater for the thick disk stars: $\Delta$[Cu/Fe] = $+$0.07 (Table~6).
This result was shown by Prochaska et al.
(2000). At [Fe/H] $< -0.8$, [Cu/Fe] falls below the solar
value, possibly in a precipitous way.
Our results generally confirm earlier results for disk
and halo stars
(Sneden et al. 1991; Mishenina et al. 2002; Bihain et al. 2004;
Simmerer et al. 2004).

\subsection{The $s$- and $r$-process elements}

Synthesis of elements beyond the iron group occurs by two neutron-capture
processes: the $s$-process occurring in AGB stars, and the $r$-process
occurring (probably) in Type II supernovae. In the solar system mix of
elements, barium is predominantly a $s$-process product and
europium an $r$-process product. Our suite of elements includes not
only Ba and Eu but also Y, Ce, and Nd. According to Burris et al.'s
(2000) resolution of the
solar elemental abundances the  contributions from the $s$-process
are  72\% for Y, 85\% for Ba, 81\% for Ce, 47\% for Nd, and merely
3\% for Eu.  Bensby et al. (2005) determined abundances
for Y, Ba, and Eu showing differences in the behavior of the
$s$- and $r$-process contributions between the  thin and thick disks.

Our abundances for Y, Ba, Ce, Nd, and Eu are based with one exception
on the lines used in Paper~I. The exception concerns Eu for which we
use the resonance line at 4129 \AA\ but in Paper~I we used the
subordinate line at 6645 \AA. The atomic data for the 4129 \AA\ line
are taken from Kurucz (1998) and the isotopic abundance ratio was
set at the solar value: $^{151}$Eu/$^{153}$Eu = 53/47.
The 4129 \AA\ line was analyzed in all stars of the present sample
and also those in Paper~I. The Eu abundances from the 4129 \AA\ and
the 6645 \AA\ lines are in good agreement for Paper~I's stars.

The trends
of the abundances of Y, Ba, Ce, Nd, and Eu with [Fe/H] (Figure~17)
show a difference between the contributions of the $s$-process
and the $r$-process to the thin and thick
disks, as found by Bensby et al. The elements -- Y, Ba, and Ce --
expected to have a dominant
$s$-process contribution are Ni-like elements. Our results for Y
agree with thick disk results from Prochaska et al. (2000).
However,
Bensby et al. (2005) found [Y/Fe] ratios showing a slow decreasing trend with
decreasing [Fe/H], but the [Y/Fe] ratios at given [Fe/H] are almost same
for both thin and thick disk stars similar to our present results. In the case
of Ba, our [Ba/Fe] ratios for thick disk stars are on average smaller by 0.1 dex
than for thin disk stars. There also may be small trend of decreasing [Ba/Fe] with
decreasing [Fe/H]. Again our and Prochaska et al. results for Ba agree,
both in trend and magnitude.
Both Bensby et al. (2005) and Allende Prieto et al. (2004) found thin disk
stars to show a possible maximum of [Ba/Fe] at [Fe/H] $\sim -0.2$, but
such feature is not present in our data.

\renewcommand{\thetable}{6}
\begin{table}
\centering
\caption{ The predicted uncertainty, $\sigma_{mod}$ and the
 $\sigma_{gau}$ resulting from a Gaussian fit to the residuals for thin and thick disk stars
are given.
}
\begin{tabular}{@{}lrrrr@{}}
\hline \hline
 \lbrack X/Fe\rbrack       & \multicolumn{2}{c}{  Thin Disk} & \multicolumn{2}{c} { Thick Disk }  \\
                           &  $\sigma_{\rm mod}$  & $\sigma_{\rm gau}$ & $\sigma_{\rm mod}$ & $\sigma_{\rm gau}$ \\
\hline
\lbrack Fe/H\rbrack   & 0.07 & ... &0.08& ...  \\
\lbrack C/Fe\rbrack   & 0.14 & 0.07&0.14& 0.09 \\
\lbrack O/Fe\rbrack   & 0.16 & 0.07&0.19& 0.07  \\
\lbrack Na/Fe\rbrack  & 0.03 & 0.04&0.05& 0.07\\
\lbrack Mg/Fe\rbrack  & 0.04 & 0.04&0.05& 0.07 \\
\lbrack Al/Fe\rbrack  & 0.04 & 0.05&0.05& 0.08\\
\lbrack Si/Fe\rbrack  & 0.05 & 0.04&0.07& 0.06\\
\lbrack Ca/Fe\rbrack  & 0.03 & 0.04&0.03& 0.06\\
\lbrack Sc/Fe\rbrack  & 0.11 & 0.05&0.11& 0.09\\
\lbrack Ti/Fe\rbrack  & 0.03 & 0.04&0.06& 0.06\\
\lbrack V/Fe\rbrack   & 0.04 & 0.04&0.06& 0.10 \\
\lbrack Cr/Fe\rbrack  & 0.02 & 0.03&0.03& 0.04\\
\lbrack Mn/Fe\rbrack  & 0.04 & 0.04&0.03& 0.04\\
\lbrack Co/Fe\rbrack  & 0.02 & 0.04&0.03& 0.06\\
\lbrack Ni/Fe\rbrack  & 0.02 & 0.03&0.04& 0.04 \\
\lbrack Cu/Fe\rbrack  & 0.02 & 0.06&0.04& 0.08 \\
\lbrack Zn/Fe\rbrack  & 0.05 & 0.06&0.10& 0.06 \\
\lbrack Y/Fe\rbrack   & 0.09 & 0.07&0.11& 0.12 \\
\lbrack Zr/Fe\lbrack  & 0.10 & 0.07&0.11& 0.12 \\
\lbrack Ba/Fe\rbrack  & 0.12 & 0.08&0.12& 0.11\\
\lbrack Ce/Fe\rbrack  & 0.11 & 0.08&0.11& 0.10 \\
\lbrack Nd/Fe\rbrack  & 0.10 & 0.07&0.11& 0.09\\
\lbrack Eu/Fe\rbrack  & 0.11 & 0.08&0.13& 0.08\\
\hline
\end{tabular}
\end{table}

Europium is  a Mg-like element.
The scatter in [Eu/Fe] at a given [Fe/H] is attributable
to the measurement errors. The large scatter in Eu results for the
thick compared to the thin disk may be attributed to large number of cooler
stars in thick disk sample.
Continuum placement at 4129~\AA\ is more difficult for the cooler stars, and
also sensitivity to the model parameters is different.
Eu results agree quite well with the Eu results
from Prochaska et al. (2000) and Bensby et al. (2005).

Neodymium which is provided about equally by the $s$- and $r$-process
in the solar mix of abundances can be imagined from Figure~17 to be
a mix of a Mg-like and a Ni-like element. We do not confirm the
trend of increasing [Nd/Fe] with decreasing [Fe/H] reported
by Allende Prieto et al. (2004) for thin disk stars.

\subsection{Cosmic scatter}

In Paper~I, we showed that the scatter of the [X/Fe] values about the
mean trend was entirely dominated by the measurement errors, i.e., there
was no hint of a cosmic scatter in the values found for the thin disk
stars. It is of value to determine if the thick disk stars betray
cosmic scatter in their [X/Fe] values.

We estimate the scatter in [X/Fe] for the thick disk
by removing the linear fit made to the data points for [Fe/H] $< -0.3$
and fitting a Gaussian to the
residuals -- see Figure~18 for [Si/Fe].
In Table~5, we list the Gaussian $\sigma_{\rm gau}$
and the expected $\sigma_{mod}$ from the congerie of measurement
errors. The same quantities from Paper~I are also tabulated.
The  comparison of  $\sigma_{gau}$ with $\sigma_{mod}$ (Figure~19)
shows that they are very similar, i.e., there is no detectable
cosmic scatter for the sampled elements for thick disk stars with
[Fe/H] $< -0.3$.
Oxygen's $\sigma_{mod}$ determined from the [O/Fe] obtained
from the O abundances based on the O\,{\sc i} 7772 \AA\ line,
after introducing an
empirical correction for departures from LTE, greatly exceeds the
measured scatter ($\sigma_{gau}$) which is apparently overestimated for these
high-excitation lines.

\section{Chemical evolution of the thick disk}

By chemical evolution is meant the run of [X/Fe] versus [Fe/H] for a suite of
elements X sampling the major sites and processes of stellar nucleosynthesis. Here,
the Mg-like elements separate thick from thin disk stars, and it is their
evolution that we address first. To reduce the observational uncertainties as much
as possible, we use the composite index [$\alpha$/Fe] from the average of the
four indices [X/Fe] for X = Mg, Si, Ca, and Ti. The individual indices are
given equal weight.
The index [$\alpha$/Fe] versus [Fe/H] is shown in Figure~20.
Two representations of the thick disk's chemical evolution are suggested by inspection
of Figure~20.

First, as proposed by Bensby et al. (2003), [$\alpha$/Fe] of the thick
disk is greater than for the thin disk for [Fe/H] $< -0.3$ but at about a [Fe/H] of
$-0.3$, the [$\alpha$/Fe] appears to tend towards thin disk values
and for [Fe/H] $> -0.2$, thick and
thin disk stars are chemically identical to within the measurement errors. Bensby et al.
speak of a `knee' linking the thick disk stars of [Fe/H] $< -0.3$ with those of
[Fe/H] $> -0.2$.  This puts all but a  few of our thick disk stars into a single relation.
The exceptions are five stars in Figure~20
with [Fe/H] $< -0.3$ with thick disk kinematics but a
thin disk [$\alpha$/Fe].

In an alternative representation proposed here, the thick disk is restricted
to [Fe/H] less than about $-0.3$, and the above five
exceptions with thin disk compositions but thick disk kinematics
are considered to form a single
relation with the thick disk stars of [Fe/H] $> -0.3$.
Thirteen of our thick disk stars from across the [Fe/H] spanned by the thin disk
form this latter
 relation which is considered distinct from the more populated
thick disk relation between [$\alpha$/Fe] and [Fe/H] $< -0.3$.
In this representation,
there is no knee connecting thick and thin disk stars.

Before commenting further on the two representations,
we remark upon the stars -- five
in one and thirteen
in the other representation -- with the kinematics of the
thick disk but abundances of the thin disk: here,
 we refer to these as the TKTA stars.

\subsection{The TKTA stars}

TKTA stars have {\it contaminated} some previous studies.
For example,
the eight thick disk stars in Mishenina et al. (2004) sample
with [Fe/H] $> -0.3$ had the abundance pattern of the thin disk. Application of
our membership criteria reduces the octet to a single thick disk star. Although
all eight have a negative $V_{LSR}$ ($-40$ to $-100$ km s$^{-1}$) outside the normal
range for thin disk stars, they with a single exception have a low $W_{LSR}$ and
so remain close to the Galactic plane, a fact noted by Mishenina et al. The
exception at [Fe/H] of $+0.12$ is a TKTA star by our definition.

Bensby et al. (2003, 2005) were the first to suggest that thin and
thick disk abundance patterns converged for [Fe/H] $> -0.2$.
According to our criteria, 17 of their
stars belong to the thick disk. One of the 17 is a TKTA star with [Fe/H] $= -0.37$.
Other potential TKTA stars are at metallicities for which thin and thick disk have the
same abundance pattern.

In our larger sample, we have five TKTA stars at a low [Fe/H] such that their
abundances clearly
separate them from the majority of the thick disk stars.
 It is unlikely that the five are mistakenly identified as
TKTA stars through a conspiracy of errors.
 The individual
[X/Fe] contributing to the [$\alpha$/Fe] each mark them out as having
thin disk abundances.
One of the five (HIP 50671) was earlier analysed by
Edvardsson et al. (1993) who obtained a similar  [Fe/H] ($-0.42$ vs our $-0.48$)
and [$\alpha$/Fe] ($0.05$ vs our $0.06$). Another (HIP 31188) was treated by
Fuhrmann (1998) whose  [Fe/H] was $-0.81$ vs our $-0.59$ and [Mg/Fe] was $+0.22$
vs our $+0.12$. If placed in Figure~9, the star would fall on an extension of
Fuhrmann's thin disk abundances and below his thick disk abundance of [Mg/Fe] of
0.4 for his [Fe/H]. It should be considered a possible TKTA star when judged
solely by Fuhrmann et al.'s abundances.
A fresh study of HIP 31188 may
be desirable but we note that our Si, Ca, and Ti abundances support our
low value of [Mg/Fe].

The set of  thirteen TKTA stars
with the additional examples from
Mishenina et al. and Bensby and colleagues  run in [Fe/H] over the entire metallicity
spread of the thin disk.
The close correspondence between the compositions of TKTA stars and
the thin disk suggest that there is a close relation between the two
groups.
Heating of the thin disk is a well known phenomena: the dispersion
of the components ($U_{LSR}$, $V_{LSR}$, $W_{LSR}$) increases with age. Nords\"{o}m et al.
(2004)- provide a recent determination of the dispersion -- age relations.
Perhaps, the  TKTA stars are thin disk stars which have been heated to resemble
the genuine thick disk stars. The old ages of the TKTA stars favour above
average heating and, perhaps, owing to the stochastic nature of the
heating processes, the TKTA stars were subjected to much greater than
normal heating. Low mass analogues of the runaway B
stars are a possibility. Against the association of the TKTA stars with the
thin disk is the fact that on average the TKTA stars appear to be
systematically a couple of Gyr older than the oldest thin disk stars (Figre
24 below).

The TKTA stars appear to be confined
to certain parts of the ($U_{\rm LSR},V_{\rm LSR},W_{\rm LSR}$)
space. Figure~21 using
[$\alpha$/Fe] to distinguish thick from thin disk stars
shows that the TKTA tag on to the low $V_{\rm LSR}$ tail
of the thin disk stars.
The TKTA stars  favour positive $U_{\rm LSR}$ over
negative values but are distributed in
$W_{LSR}$ in a similar way  to thick disk stars.
A wider exploration of ($U_{\rm LSR},V_{\rm LSR},W_{\rm LSR}$)
space may reveal some clumping of TKTA stars, an indication of
a moving group.

\subsection{The knee}

In the representation in which TKTA stars are given an  identity
apart from the thick disk stars
(i.e., thin-disk stars with special kinematics), thick disk stars
are restricted to
[Fe/H] less than about $< -0.3$ or $-0.2$.
The alternative representation proposed by Bensby et al. (2003)
introduces
a knee in the  thick disk's
[X/Fe] vs [Fe/H] for a Mg-like element such that the near-constant
[X/Fe] for [Fe/H] $< -0.3$
merges smoothly with the lower [X/Fe] values of  thin and thick disk
results for [Fe/H] $> -0.1$ or so. (The TKTA stars with
[Fe/H] $< -0.3$ still require a special interpretation in this scenario.)
 Interpretations of the chemical
evolution of the thick disk  differ for these different
representations.

The leading suggestion of a `knee' was made by Bensby and
colleagues.
Comparison with our results is  made by combining the
index [$\alpha$/Fe] from the abundances obtained by Bensby and
colleagues with our probability criteria for stellar populations.
Figure~22 shows the result where thick disk stars by our criteria
are shown by the larger filled triangles and thin disk stars by small
 open triangles.
Small filled triangles represent stars with a probability of 50 to 70\%
of belonging to the thick disk (these were assigned to the thick disk by
Bensby et al.). Compare this figure with Figure~20. Although a knee could
be drawn in Figure~22, our representation of a thick disk terminating at
about [Fe/H] of $-0.3$ and a separate thin disk relation with superposed
TKTA stars would appear to be as satisfactory a fit. Figure~22 shows
one TKTA star (HIP 3704) with [Fe/H] $< -0.3$ but three with [Fe/H] $> -0.1$.
The thick disk relation may show a slight drop in
[$\alpha$/Fe] near [Fe/H] $= -0.3$.

Bensby et al. would likely contend that the principal evidence for the knee lies
in their
results for oxygen (Bensby et al. 2004, 2005) which show a
near-linear and continuous relation for [O/Fe] vs [Fe/H] in
thick disk stars with no hint of a discontinuity at [Fe/H]
$\sim -0.3$. Existence of the knee would appear to depend, however,
on the [O/Fe] values for just two to four stars in the interval
[Fe/H] of about $-0.3$ to $0.0$. The oxygen abundances derived
from the [O\,{\sc i}] 6300 \AA\ line (corrected for a blending
Ni\,{\sc i} line) provide the clearest evidence for the
continuous relation. Abundances from the O\,{\sc i} 7772 \AA\
triplet do not offer decisive evidence in favor of a continuous
relation for the thick disk stars.
Ram\'{\i}rez et al. (2005), who performed a non-LTE analysis
of the O\,{\sc i} 7772 \AA\ triplet for a large sample
of thick disk stars (our present sample plus about an
additional 20 stars) suggest that the thick disk [O/Fe] versus [Fe/H]
relation
show no clear signs of a knee.

Marsakov \& Borkova (2005) attempt to separate chemical
evolution for thick and thin disks by discussing [Mg/Fe]
indices compiled from published values collated by
Borkova \& Marsakov (2005) who  attempt
to place the collated results for [Fe/H] and [Mg/Fe]
 from about 80 publications on common
scales. Velocity components ($U,V,W$)
are also provided.  Thick disk stars are separated from
thin disk stars by criteria that give results similar to
ours. Marsakov \& Borkova conclude that [Mg/Fe] for the
thick disk starts to decline steeply from [Mg/Fe] $\simeq +0.4$
at [Fe/H] $\simeq -0.5$ to [Mg/Fe] $\simeq +0.1$ at [Fe/H]
$= -0.3$ with a subsequent less steep decline.
This interpretation is based on abundances for 133 thick
disk stars.

The separation in [Mg/Fe] at [Fe/H] $\simeq -0.3$ between thick and
thin disk stars is not large
(say, 0.2 dex) and, therefore, close attention must be paid in
combining measurements from different authors to relative precision
and to normalization to common abundance scales for Mg and Fe.
If selection of thick disk stars is restricted to those stars
with [Mg/Fe] from two or more sources, the evidence for the
knee is greatly weakened (see  Marsakov \& Borkova's Fig. 2b) for
lack of thick disk stars with [Fe/H] $> -0.3$ (three only).
Only when stars with a single measurement of [Mg/Fe] are
included is the thick disk quite well represented across the full
[Fe/H] range (their Fig. 2c) but the clean separation between thin and thick disk stars
clearly obtained by Bensby et al. and by us is not found.
This lack of a separation must cast doubt on the conclusions
drawn by Marsakov \& Borkova.

On applying our criteria to the space velocities given in
Borkova \& Marsakov's (2005) catalogue and adopting their recommended [Mg/Fe],
we obtain the results shown in Figure~23 for stars with [Fe/H $> -2.0$.
The top panel shows  results
for thick disk stars alone; the majority of these 84 stars are drawn from the
studies we have commented upon.
This panel provides no support for a knee.
The lower panel includes the thin and halo stars drawn again using our
criteria from the catalogue.
There is a large number of stars classified as belonging
to the thin disk with a [Mg/Fe] of the thick disk.

Very recently, Brewer \& Carney (2005) offered evidence in favour of a knee.
Their sample of 23 G dwarfs contained, by their method of assigning
population membership (see Venn et al. 2004), nine thin disk and 14 thick
disk stars. The method overlooks the fact the three categories of
membership (halo, thin and thick disk) are not represented equally
in the solar neighbourhood (thin disk stars dominate the stellar
population).  On applying the method in Section 3.2 in which the
membership fractions are considered, we find that
18 of the 23 stars are thin disk stars, the other five fall in our
thin/thick disk group with a maximum probability of belonging to the
thick disk of under 50\%, and not a single star would
be assigned thick disk status by us.
Nonetheless, consistent with our results in Figure~12, Brewer \&
Carney's precise abundances delineate the thin and thick disk
trends of [$\alpha$/Fe] with [Fe/H].  There may be evidence from
stars  with [Fe/H] $\sim -0.2$ and higher of a knee, but this comes from
stars we would classify as belonging to the thin disk (see Figure~12 -- bottom
panel).

\subsection{Stellar ages}

Interpretations of the chemical evolution of the thick disk, as represented by
a plot of [X/Fe] vs [Fe/H], should take account of the ages of thick and thin
disk stars.
There is a consensus that the mean age of the thick disk is about five Gyr greater than that of
the thin disk  when isochrone ages are estimated for stars in
the solar neighbourhood. Evidence, described as tentative,
also exists  that thick disk stars describe an age-metallicity relation (Bensby et al. 2004, 2005).
Here, the age-metallicity  relation  and some connections between the
ages and other properties are presented for our sample of thick disk stars.

Determination of stellar ages was discussed in Sec. 4.5. Ages were obtained for 43 of the 95
thick disk stars. Figure~24 shows  [Fe/H] versus age for these stars together with ages for
150 thin disk stars from Paper~I. This figure gives an impression of a continuous age-metallicity
relation. For the thin disk stars this is a false impression because Paper~I did not fairly
represent stars with [Fe/H] $> -0.2$ for which the age spread is very similar to that
of the thin disk stars with [Fe/H] $< -0.2$. The thick disk stars were not knowingly
selected by [Fe/H] but by kinematic criteria.
 There is an age-metallicity
relation for the thick disk: metallicity increases to
about [Fe/H] $\sim -0.3$ in 5 Gyr from $-1.5$.
 This result is consistent with previous studies.
The maximum ages given the measurement errors are consistent with
the $\it WMAP$ age of just under 14 Gyr. As emphasized recently by
Schuster et al. (2005) from ages of halo stars the mean stellar
age, star formation began within 1 Gyr of the Big Bang.
With two exceptions, thick disk stars appear older than 8 Gyr, in agreement
with the results of Allende Prieto et al. (2005) for a sample from
the SDSS.

Those stars with well-determined ages may be used to
examine the velocity-age relations. Nordstr\"{o}m
et al. (2004, their Fig. 30) show the growth of
$U_{\rm LSR}$, $V_{\rm LSR}$, and $W_{\rm LSR}$ with
age for their sample, which is dominated by thin disk
stars. This figure serves to indicate the
outer boundaries for the growth in these
velocity components arising from dynamical
heating of thin disk stars. (We showed earlier that our ages were
in fair agreement with Nordstr\"{o}m et al.'s.)
In Figure~25, we show these boundaries
in the three panels showing velocity versus age.
Stars of Paper I fall in the main between the
broken lines. In Nordstr\"{o}m et al.'s
Figure 30, the area between the broken lines is
filled with stars to about
ages of 15 Gyr, but note that their ages systematically
differ from ours for the most-metal poor stars (see \S \ref{ages}).

Our thick disk stars are, as expected, displaced
outside the area between the broken lines in the
$V_{\rm LSR}$ versus age plot. In $W_{\rm LSR}$
versus age, the thick disk stars fall along the
broken lines, and in $U_{\rm LSR}$ versus age
the thick disk stars fall within and outside the area
between the broken lines.

\renewcommand{\thetable}{7}
\begin{table*}
\centering
\caption{ Mean [X/Fe] values ($M$) in the bin of [Fe/H] $-$0.45 to $-$0.55
and the
the coefficients ($B$) of the slope of the linear regression
fit to the runs of [X/Fe] versus [Fe/H] in the metallicity range
$-$0.3 to $-$1.0 for thin and thick disk stars. The value $\Delta$[X/Fe] in column 6
is the difference between $M_{thick}$ and $M_{thin}$.}

\begin{tabular}{@{}lrrrrr@{}}
\hline \hline
\lbrack X/Fe\rbrack   &$M_{thin}$ & B$_{thin}$  & $M_{thick}$ & $B_{thick}$ & $\Delta$[X/Fe]  \\
\hline
\lbrack C/Fe\rbrack   &0.22$\pm$0.08    & $-$0.23$\pm$0.04 & 0.35$\pm$0.08  & $-$0.23$\pm$0.06 & 0.13 \\
\lbrack O/Fe\rbrack   &0.24$\pm$0.07    & $-$0.19$\pm$0.03 & 0.36$\pm$0.19  & $-$0.25$\pm$0.05 & 0.12 \\
\lbrack Na/Fe\rbrack  &0.09$\pm$0.04    & $-$0.15$\pm$0.03 & 0.12$\pm$0.05  & $-$0.06$\pm$0.04 & 0.03 \\
\lbrack Mg/Fe\rbrack  &0.11$\pm$0.06    & $-$0.15$\pm$0.03 & 0.32$\pm$0.06  & $-$0.10$\pm$0.03 & 0.21  \\
\lbrack Al/Fe\rbrack  &0.12$\pm$0.06    & $-$0.11$\pm$0.03 & 0.30$\pm$0.09  & $-$0.01$\pm$0.06 & 0.18  \\
\lbrack Si/Fe\rbrack  &0.08$\pm$0.05    & $-$0.09$\pm$0.02 & 0.22$\pm$0.06  &$-$0.11$\pm$0.03  & 0.14 \\
\lbrack Ca/Fe\rbrack  &0.06$\pm$0.04    &$-$0.13$\pm$0.02 &  0.18$\pm$0.06  &$-$0.12$\pm$0.02  & 0.12 \\
\lbrack Sc/Fe\rbrack  &0.05$\pm$0.05    &$-$0.17$\pm$0.04 &  0.17$\pm$0.10  &$+$0.14$\pm$0.05  & 0.12 \\
\lbrack Ti/Fe\rbrack  &0.06$\pm$0.05    &$-$0.18$\pm$0.02 &  0.21$\pm$0.09  &$-$0.03$\pm$0.03  & 0.15 \\
\lbrack V/Fe\rbrack   &$-$0.01$\pm$0.04 &$-$0.10$\pm$0.03 &  0.10$\pm$0.08  &$+$0.02$\pm$0.04  & 0.11 \\
\lbrack Cr/Fe\rbrack  &$-$0.02$\pm$0.02 &$+$0.03$\pm$0.01 &  0.00$\pm$0.04  &$+$0.04$\pm$0.03  & 0.02\\
\lbrack Mn/Fe\rbrack  &$-$0.18$\pm$0.05 &$+$0.16$\pm$0.02 &$-$0.22$\pm$0.05 &$+$0.38$\pm$0.04  &$-$0.04\\
\lbrack Co/Fe\rbrack  &0.00$\pm$0.05    &$-$0.10$\pm$0.02 &  0.11$\pm$0.06  &$-$0.03$\pm$0.03  & 0.11 \\
\lbrack Ni/Fe\rbrack  &0.00$\pm$0.02    &$-$0.00$\pm$0.01 &  0.04$\pm$0.03  &$-$0.03$\pm$0.04  & 0.04\\
\lbrack Cu/Fe\rbrack  &$-$0.01$\pm$0.04 &$-$0.05$\pm$0.04 &  0.06$\pm$0.09  &$+$0.48$\pm$0.06  & 0.07\\
\lbrack Zn/Fe\rbrack  &0.09$\pm$0.07    &$-$0.14$\pm$0.07 &  0.12$\pm$0.07  &$-$0.11$\pm$0.08  & 0.03 \\
\lbrack Y/Fe\rbrack   &$-$0.03$\pm$0.1  &0.16$\pm$0.04    &  0.01$\pm$0.08  &$-$0.10$\pm$0.06  &$-$0.04\\
\lbrack Ba/Fe\rbrack  &$-$0.04$\pm$0.11 &$+$0.09$\pm$0.06 &$-$0.19$\pm$0.08 &$-$0.10$\pm$0.07  & $-$0.10 \\
\lbrack Ce/Fe\rbrack  &$-$0.01$\pm$0.10 &$+$0.00$\pm$0.05 &  0.03$\pm$0.09  &$+$0.00$\pm$0.06  & 0.04\\
\lbrack Nd/Fe\rbrack  &0.04$\pm$0.18    &$-$0.03$\pm$0.06 &  0.15$\pm$0.10  &$-$0.04$\pm$0.04  & 0.11\\
\lbrack Eu/Fe\rbrack  &0.15$\pm$0.06    &$-$0.26$\pm$0.03 &  0.38$\pm$0.11  &$-$0.20$\pm$0.09  & 0.23 \\
\hline
\end{tabular}
\end{table*}

\section{Mergers and Abundances}

Today, the origin of the thick disk is believed to lie in
the formation of the early Galaxy through mergers of smaller (proto-)
galaxies in the context of a $\Lambda$CDM universe.
Simulations of galaxy construction by mergers are
regularly reported in the literature. Reviews are written to correlate
observational and theoretical evidence on the thick disk not only of the
Galaxy but of (edge-on) spiral galaxies.
Here, we restrict remarks largely to interpretations of the
abundance differences between thick and thin disk and the
similarities between the thick disk and the halo. The
reader is referred to comprehensive reviews for
a more detailed discussion of observations and theory --
see, for example, Majewski (1993) and Freeman \&
Bland-Hawthorn (2002). Our discussion is obviously
influenced by recent commentaries by Dalcanton (2005) and
Wyse \& Gilmore (2005).

Recent detections of accretion of dwarf galaxies show that
merger is a continuing way of life for the Galaxy (e.g., Ibata et al. 1997,
Yanny et al. 2003).
Theoretical ideas about mergers in the usual $\Lambda$CDM
universe predict that the rate of mergers was much higher in
the past with a marked decline in the rate at (say) a redshift $z \sim 1$
or about 8 Gyr ago.  Coupling of the history of mergers with the
formation of thick disk stars explains why the Galactic thick disk
stars are old. One additional condition is required: the system of
thick disk stars must be created free of gas in order that
star formation is cut off and young thick disk stars are
not formed. Thin disk stars form from gas -- likely, a mixture of
pre-merger gas of the Galaxy and infall of gas following the mergers.

A defining characteristics of the thick disk is the large
vertical scale height, that is the high vertical velocity
dispersion ($\sigma_W$). Dalcanton (2005) notes that
there are three ways in which a high $\sigma_W$ may be
achieved through merging:
(A) heating  of a thin disk in a merger
(which may or may not lead to the disruption of the thin disk);
 (B) direct accretion of stars from
satellite galaxies; and
 (C) star formation in merging
gas-rich systems.  There is the real possibility that the
three ways each contributed to formation of our thick (and thin) disk.

\subsection{Scenario A}

The Galaxy develops a thin disk with on-going star formation.
Merger with one or more satellite galaxies occurs. Thin disk
stars are heated in the merger to create a thick disk of
stars but not gas. Thick disk stars are fossils from the
early years of the thin disk. This thin disk of gas is largely destroyed
but reforms following the mergers by accreting gas from the
original thin disk, the satellites, and, perhaps, the proverbial
infall of (primordial?) gas. After a hiatus, star
formation resumes in the thin disk
from gas having an initial abundance set by the compositions of
the contributors of gas and may be also by the Type Ia supernovae
from the generations of thick disk stars. Quinn, Hernquist \& Fullagar (1993)
is commonly cited as providing a description of thick disk
formation through a merger.

As pointed out by Dalcanton (2005), vertical
heating does not change angular momentum significantly. For example,
Quinn et al. (1993) show only a mild reduction of about 10 km s$^{-1}$ in
the rotational velocities of the stars after disk heating by a satellite
merger. Thus, in  this scenario it may be difficult to explain the larger
differences observed between the asymmetric drifts of the thin and thick
disks.

\subsection{Scenario B}

An alternative consequence  of galaxy growth through mergers is that
the thick disk is composed of stars accreted from satellite
galaxies.
Abadi et al. (2003a, 2003b)
describe results from a single simulation of
galaxy formation in a $\Lambda$CDM universe. Many structural and
dynamical properties of the Galaxy are satisfactorily reproduced
including a disk with well-defined thin and thick disk
components. The
majority (about 60\%) of the thick disk stars are stars captured from
satellite galaxies. This percentage increases with age; ninety
per cent of thick disk stars older than about 10 Gyr were provided
by satellite galaxies. In addition, about 15\% of the oldest
thin disk stars are from the satellites.

One of the most interesting predictions of the Abadi et al. scenario
is that thick disk stars acquired from any given satellite do not end up
at all galactocentric distances, but they rather form a ring-like
structure. Thus, different satellites may dominate the thick disk population
at different radii. However, the peak of the metallicity distribution
of thick disk stars appears to be fairly uniform between 4 and 14 kpc from
the galactic center (Allende Prieto et al. 2005).

The demonstrated uniformity of composition among thick disk
stars in the solar neighborhood (see Sec. 5.7),
and the lack of cosmic scatter, would  suggest
that these stars may not have come from a variety of satellites,
each with a different, if similar, chemical history.
It may be noted too that abundance analyses of stars in dwarf
spheroidal galaxies and the
Magellanic clouds (Hill 2004; Shetrone et al. 2003) show a variety of
[X/Fe] versus [Fe/H] relations, each unlike those of the thick disk.
However, the satellites that merged long ago may have had
abundance patterns distinct from those of the surviving dwarf
spheroidal and irregular galaxies.
Abadi et al. stress that they have only a single simulation
and, perhaps, other simulations will provide a Galaxy
with a thick disk free of a significant population
of accreted stars.
(Abadi et al. predict that one in two stars in the halo are
from satellite galaxies. Star-to-star scatter
in [X/Fe] including [$\alpha$/Fe] may exist among halo stars.)

The mergers disrupt the gas of the early thin disk.
Just prior to its disruption, a part of the  thin disk had
attained [Fe/H] $\simeq -0.3$, the maximum metallicity of the thick disk.
 After the merger, the
thin disk is  reconstituted from gas of the former thin disk and
metal-poor (presumably) gas of the satellite galaxies.
One supposes that the present thin disk began life with
[Fe/H] $\sim -0.7$, the
low metallicity bound to the present thin disk stars.
In the hiatus before
reconstitution of the thin disk, the gas is contaminated
by ejecta from Type Ia supernovae, as in scenario (C) -- see, our
discussion of Figures~26 and 27 which is also pertinent to this
scenario.

\subsection{Scenario C}

The  thick disk is formed during an early period of multiple
mergers involving gas-rich systems. A high star formation
rate is  associated with this period.
Interactions between the merging systems and the
nascent galaxy lead to a relatively hot system of
stars supported by rotation -- that is a thick disk. Subsequent to this early
period, the thin disk is formed from infalling gas.

This scenario is developed by Brook et al. (2004, 2005a, 2005b). Their
2005 paper presents abundance predictions, specifically
[$\alpha$/Fe] versus [Fe/H], for several
simulated galaxies provided from  an N-body SPH code. These
predictions (Figures 12 to 15 from Brook et al. 2005a)
 bear a remarkable resemblance to the observed
abundance patterns of the halo, thick and thin disks.

For example,  the predicted metal content of  thick disk
stars at different galactocentric distances is approximately uniform,
and so is also with distance from the Galactic plane,
in agreement with recent SDSS observations (Allende Prieto et al. 2005).
Of particular note is the clear separation in the [$\alpha$/Fe]
versus [Fe/H] relations of the thick and thin disk with each trend showing
a weak [Fe/H] dependence. In the reported simulations, the
thick disk extends in [Fe/H]  variously from about $-1.5$ to
an upper limit in the range  $-0.1$ to $-0.5$. The thick disk's
[$\alpha$/Fe] at the upper limit in two of the four reported
simulations remained separated from the thin disk's
[$\alpha$/Fe] and in the other two approached but did not merge with the
[$\alpha$/Fe] of the thin disk (i.e., the simulations do not
predict a knee). The lower [$\alpha$/Fe]  of the
thin disk results from contamination of the thin disk by ejecta from
Type Ia supernovae from thick disk binaries. The
thin disk has a metallicity spread of $-0.5$ to $+0.4$ in three of
the four simulations and $-0.9$ to $-0.1$ in the fourth.
Halo stars extend to [Fe/H] of about $-0.9$ to $-0.6$ with a
[$\alpha$/Fe] very similar to that of the thick disk in
three simulations and about $0.2$ dex larger in the fourth
simulation.

Our data on thick and thin disk abundances may be used to
characterize the pollution of the gas by SN Ia ejecta (and
AGB stars).
Ages of thin and thick disk stars show (Figure 24) that
several Gyr elapsed between the formation of the first thick
and the first thin disk stars and, thus, time aplenty for
SN Ia's (and AGB stars)
 to cause pollution. Compositions of thick disk stars up to
the most metal-rich (in our interpretation)
at [Fe/H] $= -0.3$
are dominated by contributions
from
SN II. Addition of SN Ia ejecta to gas, as recognised in a
classic paper by Tinsley (1979), increases the abundance ratio of
the iron-group elements to $\alpha$-elements.

This episode in chemical evolution of the Galactic disk  is
represented schematically by Figure~26 in which we plot
[X/$\alpha$] versus [$\alpha$/H]. The $\alpha$ abundances
are dominated by SN II products but not without a
SN Ia contribution, especially for the heavier $\alpha$-elements.
Oxygen would probably provide a cleaner measure of SN II
products, albeit with a bias to the SN II from the
most massive stars, but, at present, we do not have non-LTE
corrected O abundances for all of our stars.
Evolution of the thick disk is represented by the
track A to B; pristine gas is contaminated by SN II ejecta
raising [$\alpha$/H] and, in this example, also raising
[X/$\alpha$];  X is here an element whose yield from SN II is
metallicity-dependent. For elements whose yield from SN II is
independent of metallicity, the track A to B will be parallel
to the $x$-axis.
Stars formed as the gas composition evolves from
A to B become today's thick disk stars.

 Star formation ceases as a thin disk forms from the gas of the
satellites that gave the thick disk stars and infalling
more metal-poor gas.
On the assumption that the
thick disk gas  of the composition corresponding to point B
is diluted with very metal-poor gas, the thin disk will initially
have the composition corresponding to point C.
Following the period of active merger, there will be a hiatus
before star formation in the thin disk resumes.
In and following the interval of
very low star formation, SN Ia's pollute the gas. Pollution prior to
resumption of star formation is represented by the track C to D;
in this example, the SN Ia's products include noticeable
amounts of X but not of the $\alpha$-elements. (If $\alpha$-elements
are produced, the track C to D is slanted.) Finally, the composition of
the gas (and stars) of the new thin disk evolves along the track
D to E in response to contributions by the SN II arising from
massive stars formed in the new thin disk, SN Ia in the
thick disk and later of the thin disk, and AGB stars of the thick and thin
disks.

Evolution from A to E will depend on the
element X, as we illustrate in Figure 27, where
derived abundances for four representative elements --
Mn, Fe, Ni, and Eu --
are shown with a track A to E superimposed with
point B placed at [$\alpha$/H] $ = -0.1$, and
C placed, somewhat arbitrarily, at [$\alpha$/H] $= -0.6$.
The increase in [Mn/$\alpha$] along A to B is attributed to
a metallicity-dependent Mn yield from SN II (McWilliam
et al. 2003). This is supported by the observation that the
track C to D for Mn is short.
Iron offers a contrasting track: along the thick disk segment, A to B,
[Fe/$\alpha$] is essentially constant indicating that the
relative yields of Fe and $\alpha$-elements are metallicity
independent; the large jump from C to D shows, as anticipated,
that Fe is a principal product of SN Ia's; and the slope to the track D to E
reflects the combined contributions of SN II and SN Ia
to the continuing evolution of the thin disk.
Nickel's behaviour is similar to iron's but the increment in
[X/$\alpha$] from C to D is less for Ni than for Fe.
For Eu, [Eu/$\alpha$] is approximately constant along the
track A to B indicating a lack of a metallicity dependence of the
Eu to $\alpha$-element yields from SN II. The track C to D shows
a decrease in [Eu/$\alpha$] resulting from $\alpha$-element
production by SN Ia with the absence of Eu production. The track
D to E is controlled by the yields from SN II  and SN Ia.
In summary, chemical evolution of the disk as depicted by
Figure 26 offers a qualitative account of the compiled
data on the abundances of thin and thick disk stars.

This scenario for the formation of the thick disk
offers a fine account of the observed abundance pattern
of the Galaxy's halo, thick and thin disks. Brook et al.
describe how their simulations fit other observed
characteristics of the Galaxy and external disk
galaxies.
They also remark that accretion of
stars from satellites, after formation of the thin disk, may contribute to
the population of the thick disk  and halo, i.e., the first
scenario (above) may have a role to play.

A potential problem with this scenario,
pointed out by Dalcanton (2005), is that it will be the same
gas from the merging blocks which will give birth to
the thick disk stars first and to the thin disk population later.
This situation could make it difficult to have quite different scale lengths
for the two disks. However, as we discuss above, in order to match
the observed abundances of thin and thick disk stars, a significant
contribution of metal-poor gas must be made available before star formation
begins in the thin disk. Only in this way, the last stars formed in the
thick disk can have higher metal abundances than thin disk stars formed
at a later time. In the models reported by Brook et al., the scale
length of the thin disk, we note, is significantly different from
that of the thick disk (a ratio of 1.6), but with the thick disk
being more compact than the thin disk, which is opposite to the
behaviour observed in most galaxies, including the Milky Way.

\section{Concluding remarks}

On the basis of the  available abundance data on the
thin and thick disks and of the published simulations of disk
formation through mergers in a $\Lambda$CDM universe, scenario
(C) appears to be a plausible leading explanation for the origin
of the thick and thin disks. In terms of a continued
exploration of the observational frontiers, there is, for example, a
need for (i) a detailed abundance analysis of stars apparently
attributable to an extension of the thick disk to metallicities
below [Fe/H] of $-$1, i.e., the so-called metal weak thick disk,
which comprises only one per cent of the
thick disk (Martin \& Morrison 1998); (ii) a more complete
investigation of the four dimensional space (U,V,W,[Fe/H]), e.g.,
the pursuit of stars at low $W_{LSR}$ with [Fe/H] less than about
$-0.7$, stars, which, if present, would be assigned to the thin disk;
(iii) analysis of a larger  sample of stars with [Fe/H] greater than
about $-$0.3 and well-determined kinematics is needed to confirm or
deny the presence of the knee in the thick disk Mg-like
abundances connecting to the thin disk abundances for the
most metal-rich stars; (iv) a larger sample of thick disk stars is
needed to determine radial and vertical gradients in compositions of
thick disk stars. The vertical gradient, if any, appears to be very shallow
(e.g; Bensby et al. 2005, Allende Prieto et al. 2005).

The realisation that thick and thin disk stars of the same
[Fe/H] differ in composition and the strong suggestion
that thick and thin disk stars span overlapping but
distinctly different ranges in [Fe/H] (see, for example,
Schuster et al. 2005) has consequences for the cottage
industry providing models of chemical evolution
of the Galactic disk, especially for models of the solar
neighbourhood.
The industry standard supposes that
chemical evolution as
represented by a plot of [X/Y] vs [Y/H], where
Y is traditionally taken to Fe or sometimes O, is a continuous
process from the halo to the disk, i.e., initially
metal-free gas experiences star formation leading to
the halo stars, collapse of gas to a disk with
enrichment from stellar
nucleosynthesis and possibly continuing infall of gas leads to
a steady continous chemical evolution. No account is taken  of
the fact that the disk has the two components - thin and thick -
from different origins and most probably covering different
metallicity ranges. The time has come to change the industry standard!

\section{Acknowledgments}

We thank Jocelyn Tomkin, Kameswara Rao, and Gajendra Pandey
for many spirited and useful discussions.
This research has been supported in part
by  the Robert A. Welch Foundation of Houston, Texas.
This research has made use of the SIMBAD
data base, operated at CDS, Strasbourg, France, and the NASA ADS, USA.

{}

\renewcommand{\thetable}{2}{\arabic{table}}

\begin{table*}
\begin{center}
\caption{Atmospheric parameters and kinematic data for the programme stars. The columns 1-11
are self explanatory. The value in column 5 is the heliocentric radial velocity($R_{\rm v}$).
Errors for the values in columns 5-8 are discussed in the text.
See text for explanatory note about the probability (\%P) in column 12.} \end{center}
\begin{tabular}{rccrrrrrrrrcl}
\hline \hline
Star   &  [Fe/H] & $T_{\rm eff}$ & log $g$ & $R_{\rm v}$  & $U_{\rm LSR}$ & $V_{\rm LSR}$ & $W_{\rm LSR}$ & $R_{\rm m}$ & $e$ &
$Z_{\rm max}$ & $\tau_{9}$ & $\%$P \\
HIP     &        &    (K)    &(cm s$^{-2}$)        & \multicolumn{4}{c}{(km s$^{-1}$)}& (kpc)  & & (kpc)  & (Gyrs) \\
\hline
\multicolumn{12}{c}{Thick Disk} \\
\hline
3086&    -0.23&      5697&      4.14&  -29.2 &   -159&     -50&      56&      8.55&     0.51&     0.70&      7.9$^{+2.4}_{-1.8}$ &      95$\pm$1 \\
3185&    -0.65&      5344&      3.81&  -45.3   &-37&     -84&      45&      6.20&     0.40&     0.43&      11.3$^{+2.7}_{-2.2}$ &      87$\pm$0 \\
3441&    -0.50&      5723&      4.50&  -108.3   & 89&     -66&      86&      7.08&     0.39&      1.13&       ...     &      96.$\pm$0 \\
4039&    -1.24&      5803&      3.92&  -5.3   &-116&     -92&      46&      6.75&     0.52&     0.47&      13.0$^{+2.6}_{-2.1}$ &      95$\pm$1 \\
4544&    -0.87&      5832&      4.51& -112.2    &-33&     -133&     -74&      5.35&     0.61&     0.81&       ...     &      90$\pm$6 \\
\hline
\end{tabular}
\end{table*}
\begin{table*}
\begin{center}
\contcaption{} \end{center}
\begin{tabular}{rccrrrrrrrrcl}
\hline \hline
Star   & [Fe/H] & $T_{\rm eff}$ & log $g$ &$R_{\rm v}$  &$U_{\rm LSR}$ & $V_{\rm LSR}$ & $W_{\rm LSR}$ & $R_{\rm m}$ & $e$ &
$Z_{\rm max}$ & log $\tau_{9}$ & $\%$P \\
HIP     &       & (K)    &(cm s$^{-2}$)        &\multicolumn{4}{c}{ (km s$^{-1}$)} & (kpc)  & & (kpc)  & (
yrs) & \\
\hline
71819&   -0.14&      5825&      4.35&  -18.1   &-65&     -101&      38&      6.01&     0.50&     0.33&  ...&      94 $\pm$ 0 \\
72407&   -0.54&      5567&      4.56& 1.5    &-88&     -60&      48&      7.10&     0.38&     0.47&  ...&      81 $\pm$ 6 \\
72803&   -0.73&      5159&      3.57&  -98.9   &-98&     -85&     -45&      6.65&     0.47&     0.40&      9.6$^{+5.0}_{-3.3}$&      94 $\pm$1 \\
74033&   -0.85&      5574&      3.95&  -59.9   &-104&     -120&      33&      5.99&     0.62&     0.28&      13.8$^{+2.2}_{-1.9}$&      94 $\pm$2 \\
74067&   -0.75&      5572&      4.63& -61.1    &-20&     -60&     -61&      6.72&     0.27&     0.62&  ...&      91 $\pm$ 3 \\
81748&   -0.01&      5559&      4.29&  -76.0   &-4&     -65&     -56&      6.56&     0.29&     0.55&  ...&      86 $\pm$ 4 \\
\hline
\end{tabular}
\end{table*}
\begin{table*}
\begin{center}
\contcaption{} \end{center}
\begin{tabular}{rccrrrrrrrrcl}
\hline \hline
Star   & [Fe/H] & $T_{\rm eff}$ & log $g$ &$R_{\rm v}$  &$U_{\rm LSR}$ & $V_{\rm LSR}$ & $W_{\rm LSR}$ & $R_{\rm m}$ & $e$ &
$Z_{\rm max}$ & log $\tau_{9}$ & $\%$P \\
HIP     &       & (K)    &(cm s$^{-2}$)        & \multicolumn{4}{c}{ (km s$^{-1}$)}& (kpc)  & & (kpc)  & (
yrs) & \\
\hline
\hline\\
\multicolumn{12}{c}{Halo} \\
\hline\\
3026&     -1.04&      6060&      4.47& -50.2    & 146&     -231&     -33&      166.&     0.99&      57.2&  ...&      97 $\pm$ 2 \\
10449&    -0.87&      5566&      4.64&  27.9   &-196&     -196&      65&      6.46&     0.94&     0.77&  ...&      75 $\pm$ 0 \\
12294&    -0.95&      6069&      4.58&  54.6   &-188&     -149&      69&      6.97&     0.80&     0.81&  ...&      99 $\pm$ 3 \\
\hline
\end{tabular}
\end{table*}
\begin{table*}
\begin{center}
\contcaption{} \end{center}
\begin{tabular}{rccrrrrrrrrcl}
\hline \hline
Star   & [Fe/H] & $T_{\rm eff}$ & log $g$ &$R_{\rm v}$  &$U_{\rm LSR}$ & $V_{\rm LSR}$ & $W_{\rm LSR}$ & $R_{\rm m}$ & $e$ &
$Z_{\rm max}$ & log $\tau_{9}$ & $\%$P \\
HIP     &       & (K)    &(cm s$^{-2}$)        & \multicolumn{4}{c}{ (km s$^{-1}$)}& (kpc)  & & (kpc)  & (
yrs) & \\

\hline
\hline\\
\multicolumn{12}{c}{Thick Disk/Halo} \\
\hline\\
7452&     -1.53&      5735&      4.59& -151.2    & 34&     -193&     -46&      4.45&     0.90&     0.38&  ...&      97 $\pm$ 36 \\
14594&    -1.88&      5522&      4.55&  -140.8   & 167&     -118&     -59&      7.00&     0.69&     0.72&           ...&      73 $\pm$ 9 \\
20298&    -0.84&      5547&      4.65&   36.6  &-94&     -133&      81&      5.64&     0.66&      1.02&           ...&      79 $\pm$ 13 \\
\hline
\end{tabular}
\end{table*}

\renewcommand{\thetable}{5}{\arabic{table}}
\begin{table*}
\caption{Abundance ratios [X/Fe] for elements from C to V for the programme stars. Note that the
O abundances in column 4 are not corrected for NLTE effect}
\begin{tabular}{rrcrcrrrrrrr}
\hline \hline
HIP&   [Fe1/H]&      C&       O&             Na&        Mg&        Al&        Si&       Ca&       Sc&        Ti&       V \\
\hline
\multicolumn{12}{c}{Thick Disk} \\
\hline
      3086&   -0.23&   0.14&    0.34&     0.08&      0.15&      0.17&     0.09&           0.04&   -0.07&     0.04&       0.0\\
      3185&   -0.65&   0.28&    0.51&          0.19&      0.40&      0.34&      0.25&            0.21&     0.19&      0.27&     0.04\\
      3441&   -0.50&   0.35&    0.56&         0.06&      0.32&      0.24&      0.18&           0.10&     0.24&      0.19&     0.06\\
      4039&   -1.24&    ....&    0.71&      0.22&      0.30&     -0.12&      0.15&            0.29&    -0.13&      0.24&   ...\\
      4544&   -0.87&   0.26&    0.49&     0.09&      0.39&      0.24&      0.21&            0.11&     0.12&      0.22&      0.24\\
      5122&   -0.62&   0.44&    0.56&          0.11&      0.26&      0.30&      0.20&            0.12&     0.11&      0.18&      0.12\\
\hline
\end{tabular}
\end{table*}

\begin{table*}
\contcaption{}
\begin{tabular}{rrcrcrrrrrrr}
\hline \hline
HIP&   [Fe1/H]&      C&       O&             Na&        Mg&        Al&        Si&                  Ca&       Sc&        Ti&       V \\
\hline
     84862&   -0.41&   0.40&    0.61&         0.07&      0.35&      0.28&      0.22&            0.16&     0.23&      0.18&     0.06\\
     85373&   -0.82&    ...&    0.42&    0.01&      0.18&      0.26&      0.34&       0.05&     0.30&      0.15&      0.19\\
     85378&   -0.51&   0.27&    0.57&          0.12&      0.34&      0.35&      0.21&            0.23&     0.12&      0.23&    -0.01\\
     85757&   -0.70&   0.40&    0.72&      0.14&      0.36&      0.37&      0.29&            0.23&    0.08&      0.23&      0.11\\
     86013&   -0.70&   0.49&    0.72&     0.07&      0.31&      0.16&      0.21&        0.14&     0.18&      0.14&     0.07\\
     86830&   -0.59&   0.43&    0.69&      0.17&      0.35&      0.35&      0.30&            0.23&    0.07&      0.21&      0.14\\
     87089&   -0.30&   0.37&    0.47&     -0.12&   ...&     0.06&      0.14&       0.10&     0.16&      0.26&      0.24\\
\hline
\end{tabular}
\end{table*}

\begin{table*}
\contcaption{}
\begin{tabular}{rrcrcrrrrrrr}
\hline \hline
HIP&   [Fe1/H]&      C&       O&            Na&        Mg&        Al&        Si&                    Ca&       Sc&        Ti&       V \\
\hline
     58229&   -0.83&    ...&    0.67&     0.06&      0.18&      0.13&     0.10&       0.07&   -0.03&     0.10&   ...\\
     62882&   -0.98&   0.200&    0.80&    -0.02&      0.26&      0.12&      0.27&            0.19&    -0.12&     0.08&     0.10\\
     73385&    -1.46&    ....&     1.0&   ...&   ...&       1.6&      0.49&            0.32&      0.0&      0.33&       0.0\\
     78640&    -1.34&    ....&    0.78&   ...&      0.29&   ...&      0.23&            0.26&    -0.18&      0.26&      0.22\\
     80837&   -0.80&   0.46&    0.72&      0.14&      0.36&      0.26&      0.24&            0.16&     0.17&      0.21&      0.12\\
     86321&   -0.87&   0.30&    0.55&     0.02&      0.15&     0.07&      0.15&        0.13&     0.13&      0.12&     0.06\\
     86431&   -0.54&   0.39&    0.57&     0.05&      0.24&      0.19&      0.15&           0.07&     0.15&     0.08&     0.04\\
\hline
\end{tabular}
\end{table*}

\renewcommand{\thetable}{6}{\arabic{table}}
\begin{table*}
\caption{Abundance ratios [X/Fe] for elements from V to Eu for the programme stars}
\begin{tabular}{rrrrrrrrrrrrrrr}
\hline \hline
       HIP&   [Fe/H]&       Cr&       Mn&       Co&       Ni&       Cu&       Zn&         Y&       Ba&        Ce&       Nd&        Eu \\
\hline
\multicolumn{12}{c}{Thick Disk} \\
\hline\\
      3086&    -0.23&   -0.04&    -0.18&    0.02&   -0.03&   -0.06&     0.14&     -0.16&    -0.15&     0.04&      0.14         &0.23\\
      3185&    -0.65&    0.08&    -0.19&     0.12&    0.04&   -0.09&     0.14&     0.04&   -0.06&      0.12&      0.34         &0.60\\
      3441&    -0.50&   -0.03&    -0.32&    0.08&    0.05&   -0.08&     0.23&     0.04&   -0.09&      0.18&      0.26         &0.42\\
      4039&     -1.20&    0.07&    -0.51&     0.28&   -0.08&    -0.48&    0.08&     0.05&     0.30&   ...&   ...         &0.88\\
      4544&    -0.87&   -0.09&    -0.40&     0.13&      0.0&   -0.05&    0.08&    -0.02&    -0.14&   ...&      0.58         &0.53\\
\hline
\end{tabular}
\end{table*}

\begin{table*}
\contcaption{}
\begin{tabular}{rrrrrrrrrrrrrrr}
\hline \hline
HIP & [Fe/H] & Cr  & Mn & Co & Ni & Cu & Zn & Y &  Ba & Ce & Nd & Eu \\
\hline
     84862&    -0.41&   -0.03&    -0.24&     0.12&    0.04&    0.02&     0.26&     0.10&    -0.12&     0.06&      0.32         &0.45\\
     85373&    -0.82&    0.04&    -0.30&     0.23&    0.08&    -0.2&    0.10&     0.05&    -0.31&   ...&      0.63         &0.71\\
     85378&    -0.51&   -0.04&    -0.24&    0.09&    0.04&   -0.02&    0.09&    -0.01&    -0.15&     0.08&      0.22         &0.36\\
     85757&    -0.70&   -0.03&    -0.29&    0.10&    0.03&   -0.02&    0.10&     0.03&    -0.17&     -0.10&      0.14         &0.09\\
     86013&    -0.70&    -0.11&    -0.32&    0.08&   -0.03&    -0.13&     0.11&     0.02&    -0.13&    -0.02&      0.30         &0.35\\
     86830&    -0.59&   0.01&    -0.22&     0.14&    0.05& 0.00&    0.04&    -0.09&    -0.27&    -0.07&     0.10         &0.25\\
     87089&    -0.30&   -0.03&   -0.05&     0.12&    0.06&    0.06&    0.04&     0.05&    -0.10&      0.32&      0.28         &0.46\\
\hline
\end{tabular}
\end{table*}

\begin{table*}
\contcaption{}
\begin{tabular}{rrrrrrrrrrrrrrr}
\hline \hline
HIP & [Fe/H] & Cr  & Mn & Co & Ni & Cu & Zn & Y &  Ba & Ce & Nd & Eu \\
\hline
     58229&    -0.83&    -0.35&    -0.40&     0.26&   -0.05&    -0.44&    0.02&     0.03&    -0.13&   ...&   ...         &...\\
     62882&    -0.98&   -0.07&    -0.61&    -0.15&   -0.08&    -0.55&     0.12&     0.08&   -0.06&     0.07&      0.10         &0.12\\
     73385&     -1.46&     0.18&    -0.39&  ...&    0.07&    -0.46&     0.20&      0.20&    0.07&      0.27&   ...         &0.31\\
     78640&     -1.34&    ...&    -0.34&     0.41&   -0.03&  ...&   -0.07&    -0.07&    -0.28&   ...&      0.14         &0.06\\
     80837&    -0.80&    0.05&    -0.27&     0.22&    0.04&   0.01&     0.15&     0.06&    -0.13&    -0.05&      0.14         &0.29\\
     86321&    -0.87&   -0.07&    -0.34&     0.20&   -0.04&    -0.43&    0.08&     0.07&    0.03&   ...&   ...         &0.36\\
     86431&    -0.54&   -0.08&    -0.23&    0.05&   -0.05&   -0.03&    0.07&   0.00&   -0.07&     0.03&      0.33         &0.30\\
\hline
\end{tabular}
\end{table*}

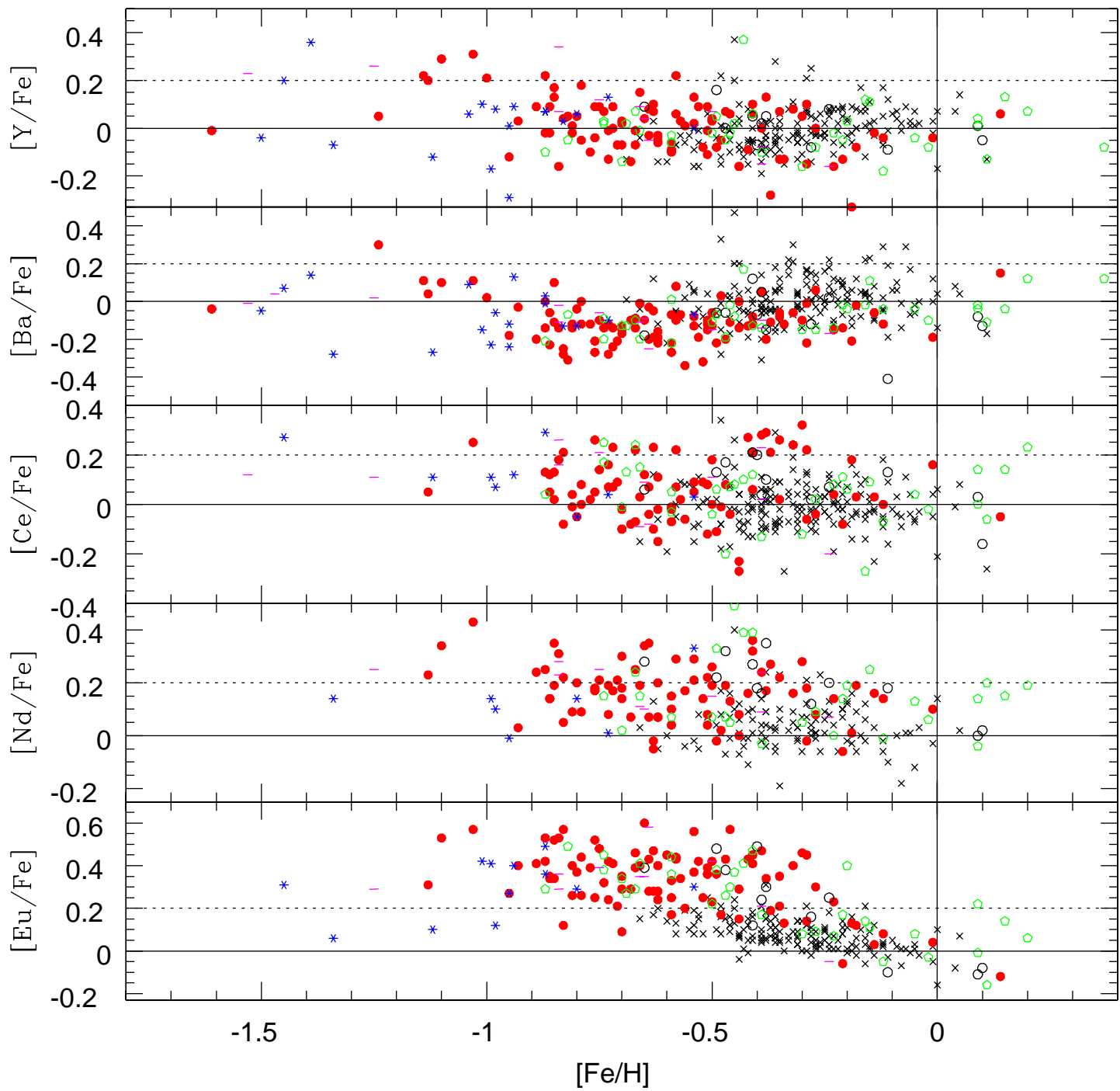
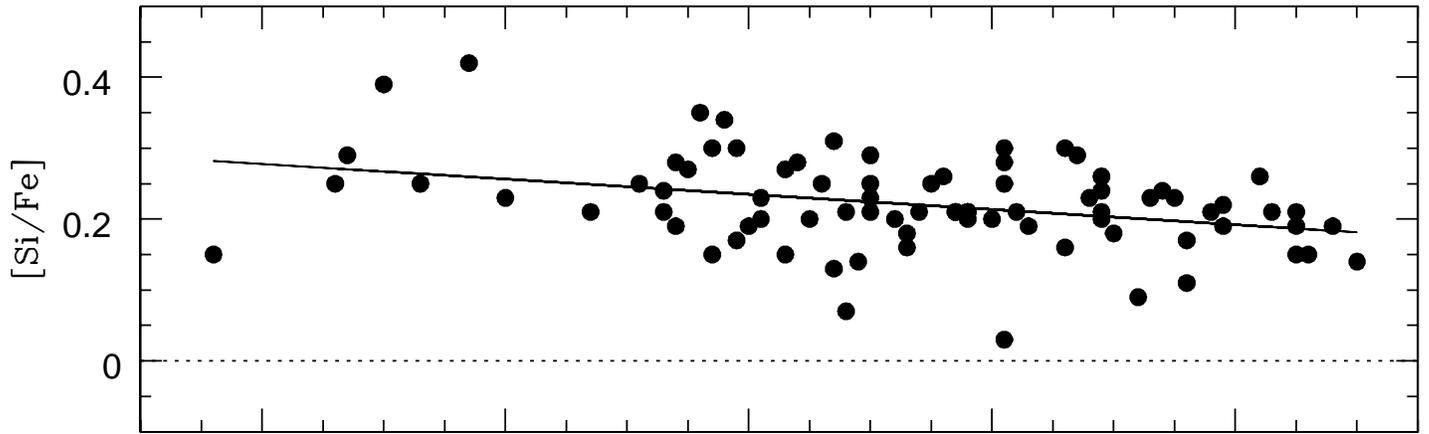
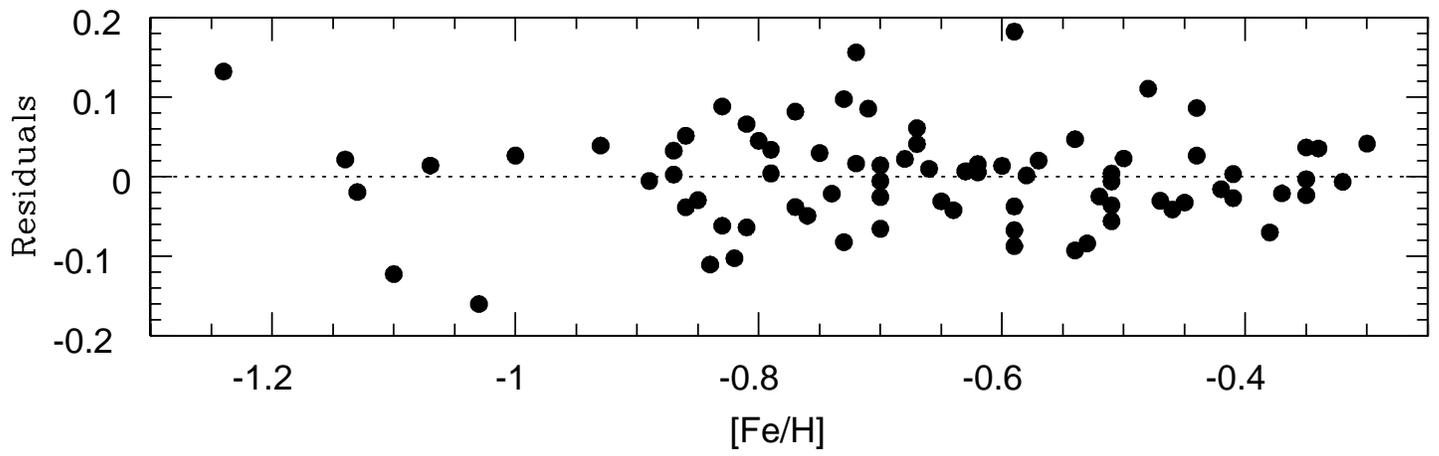
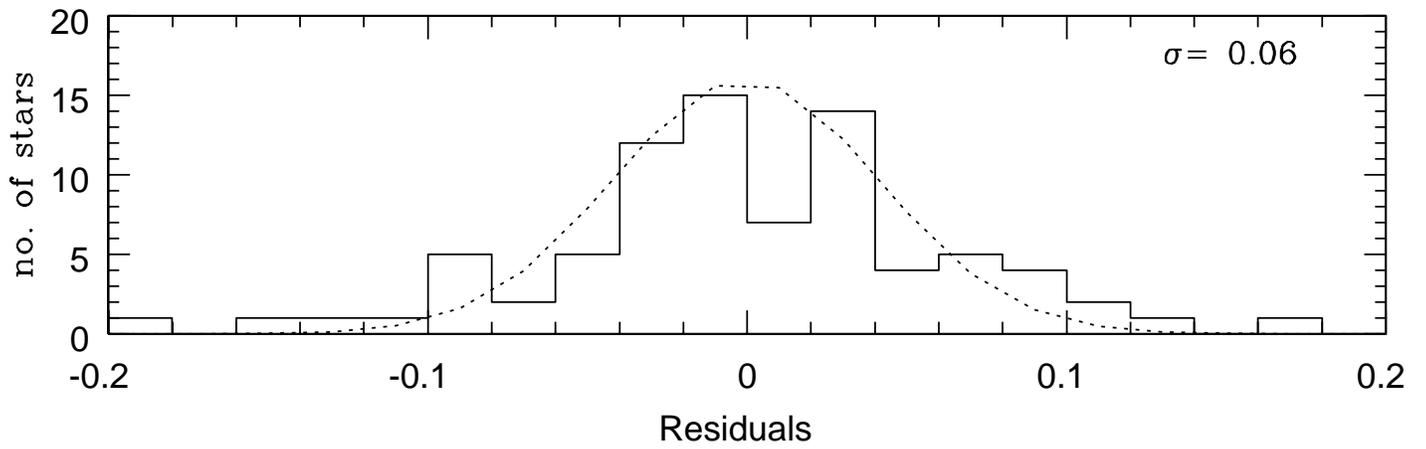
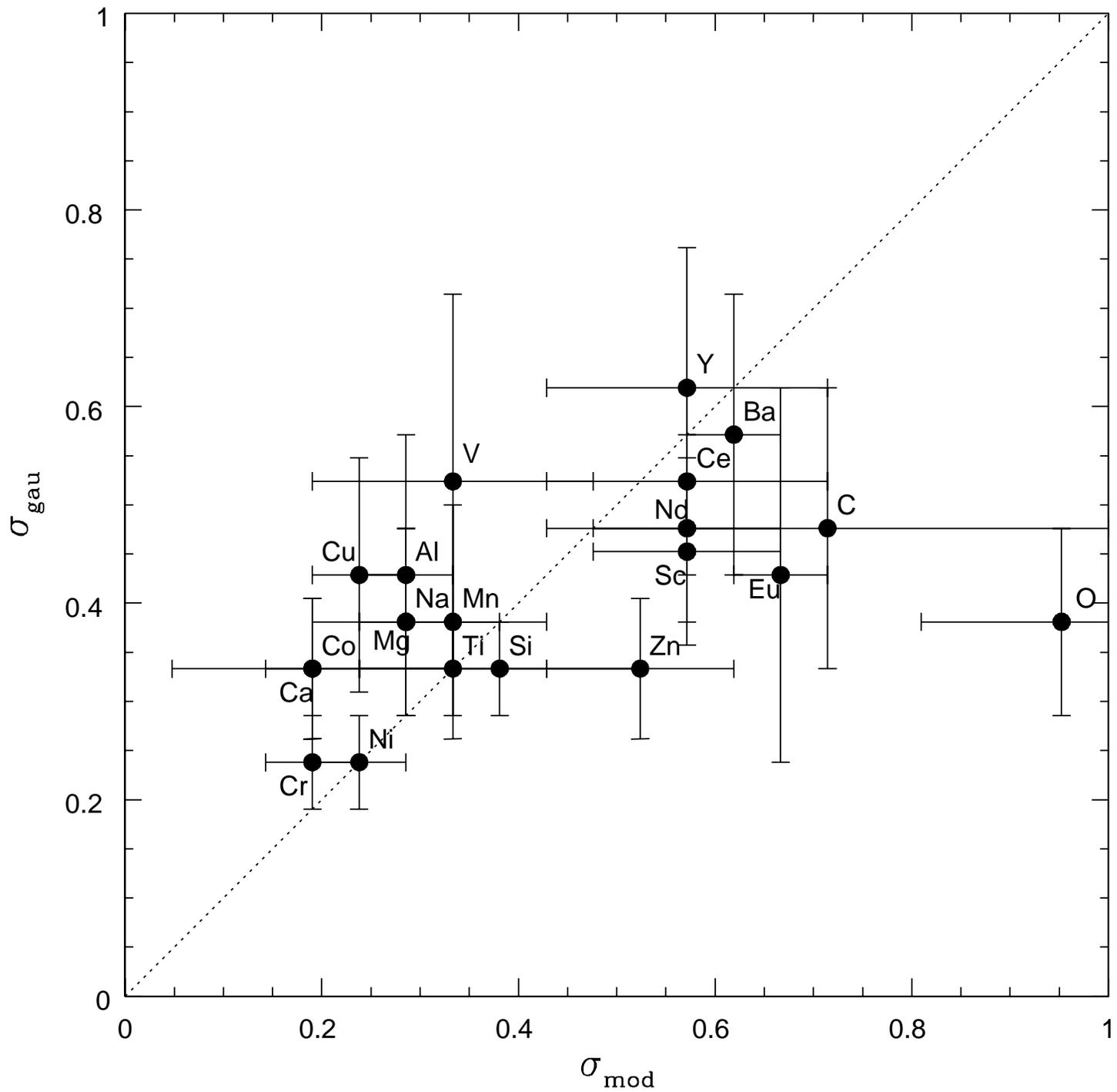
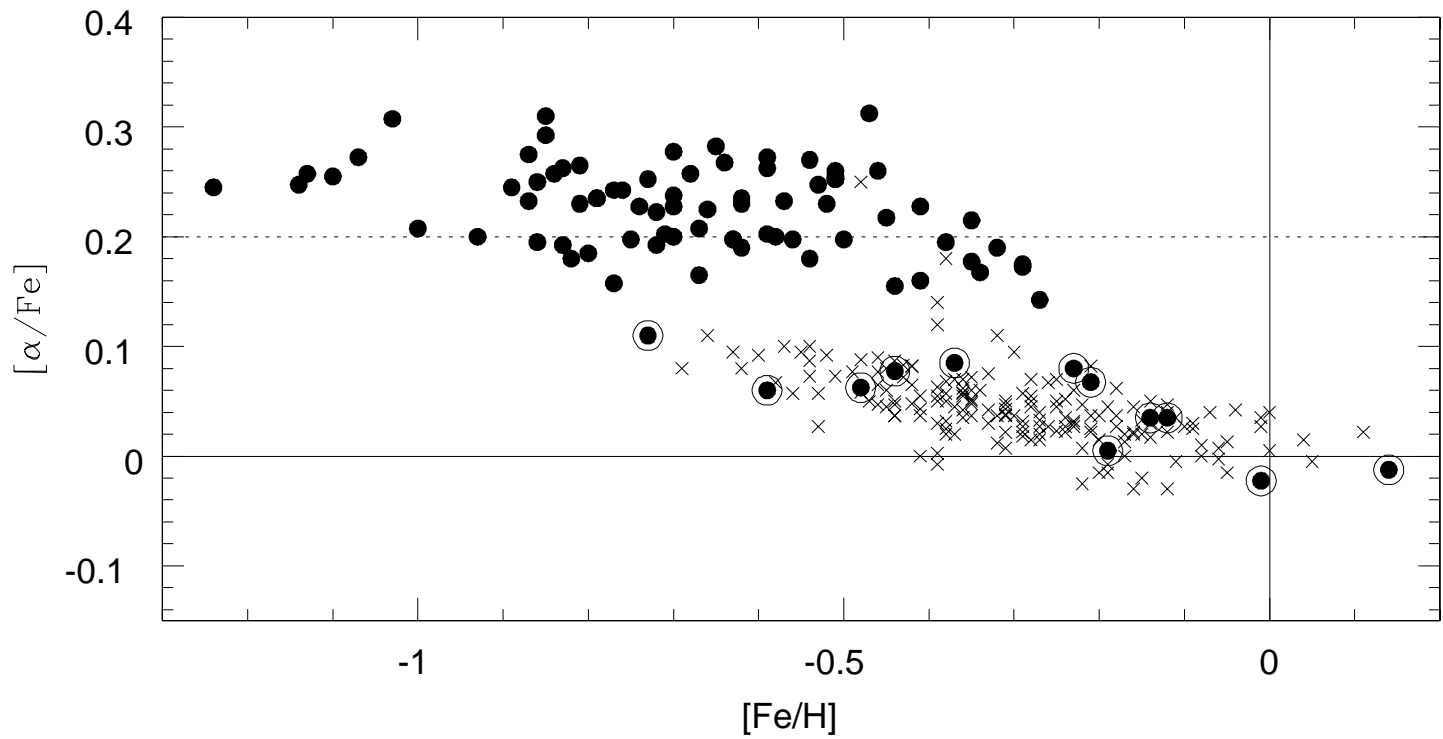
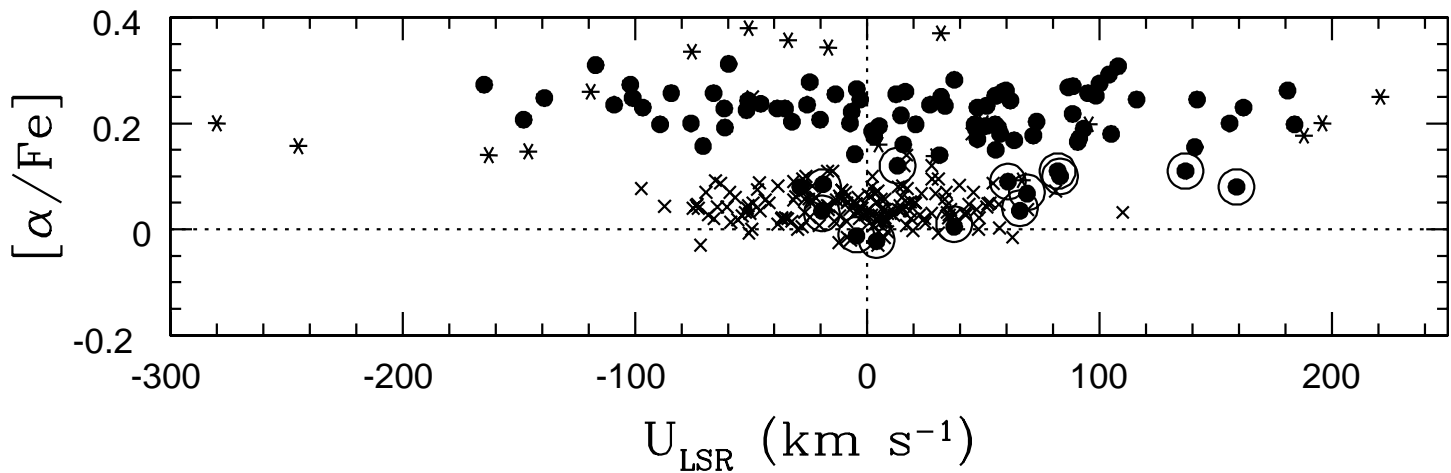
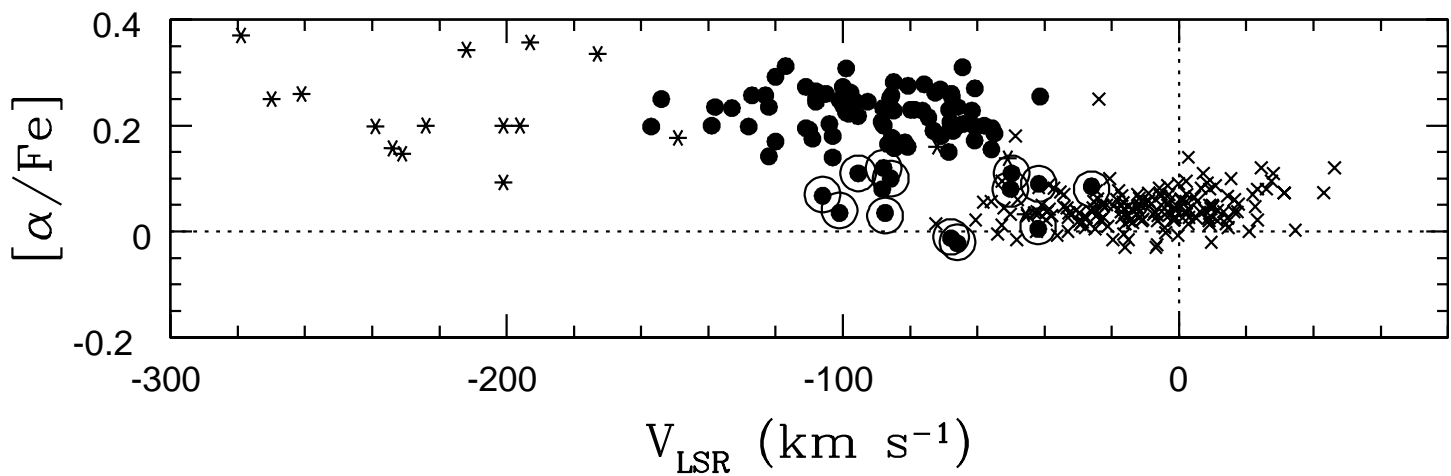
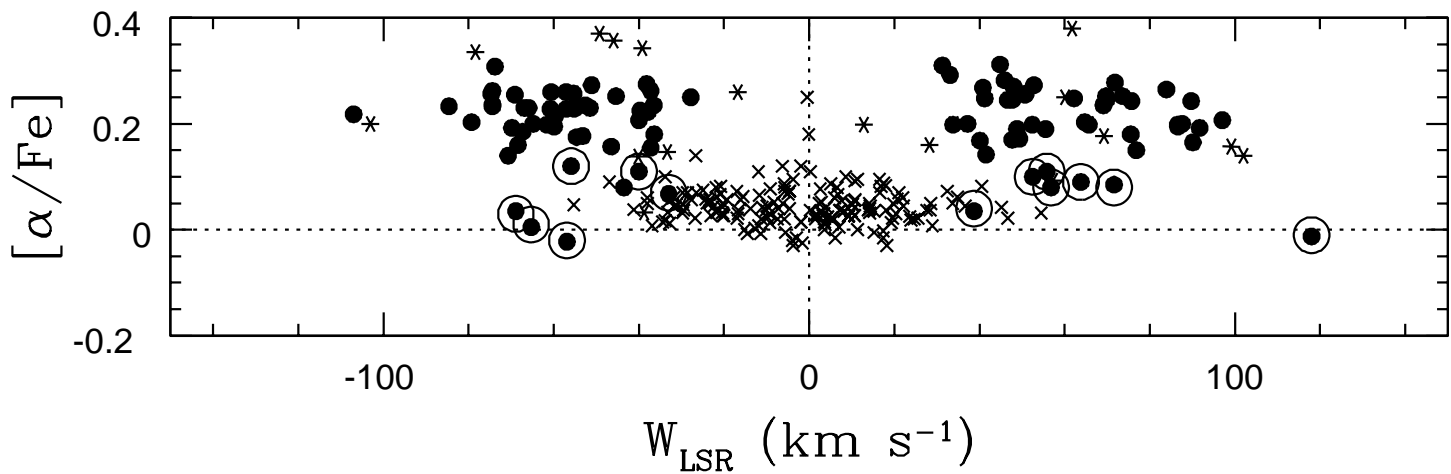
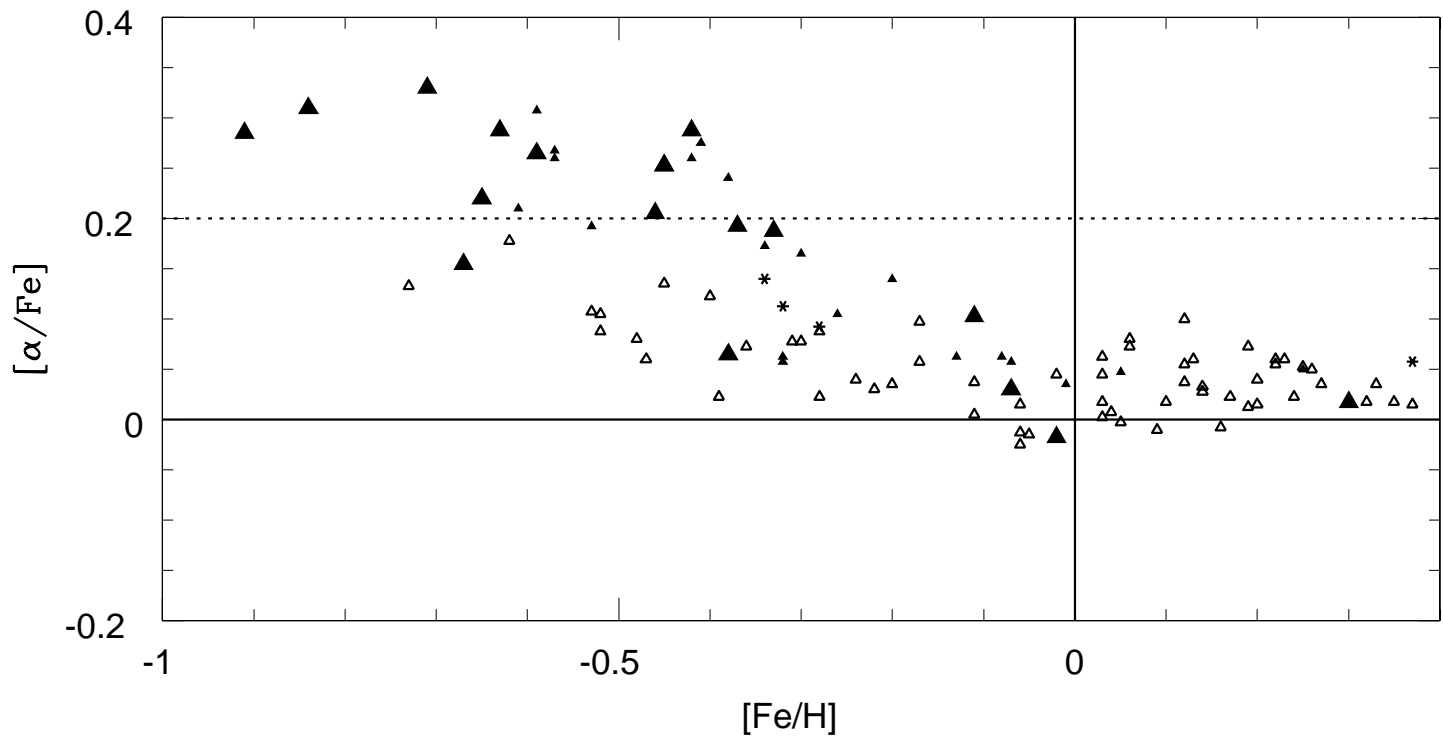
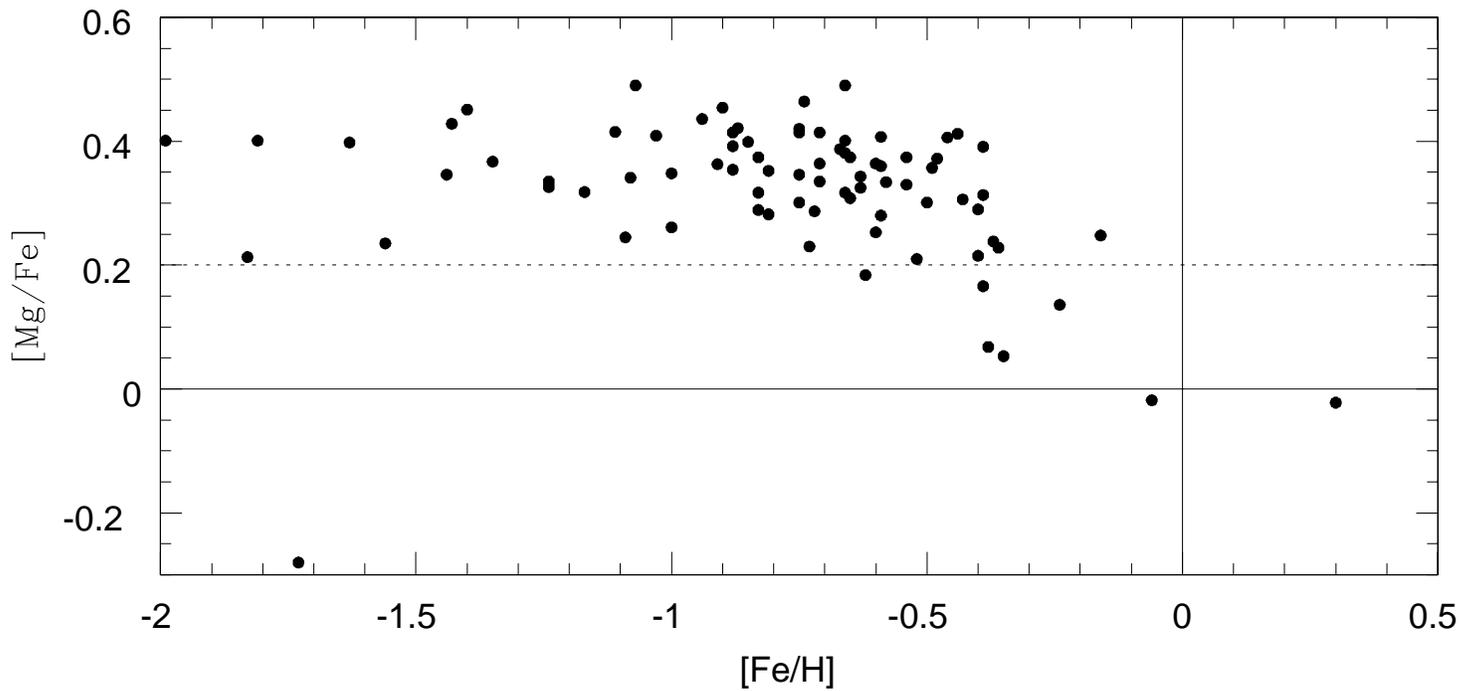
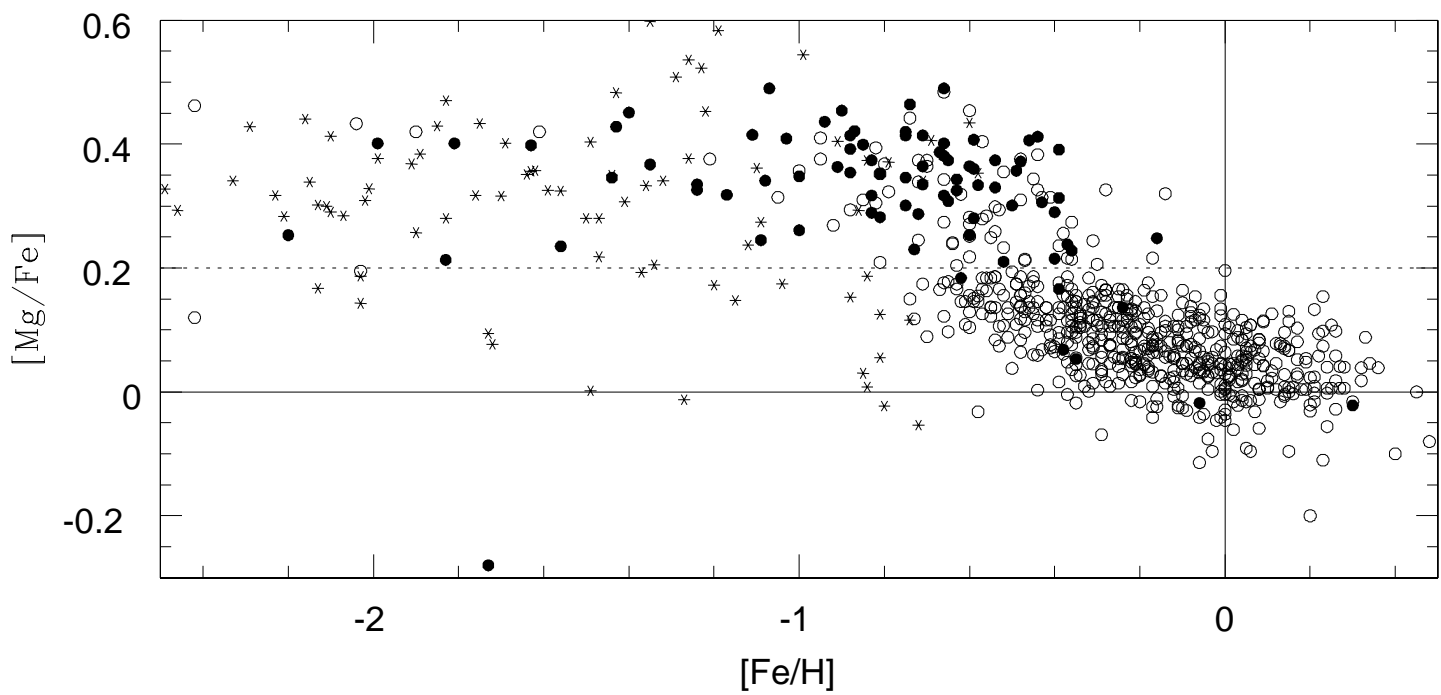
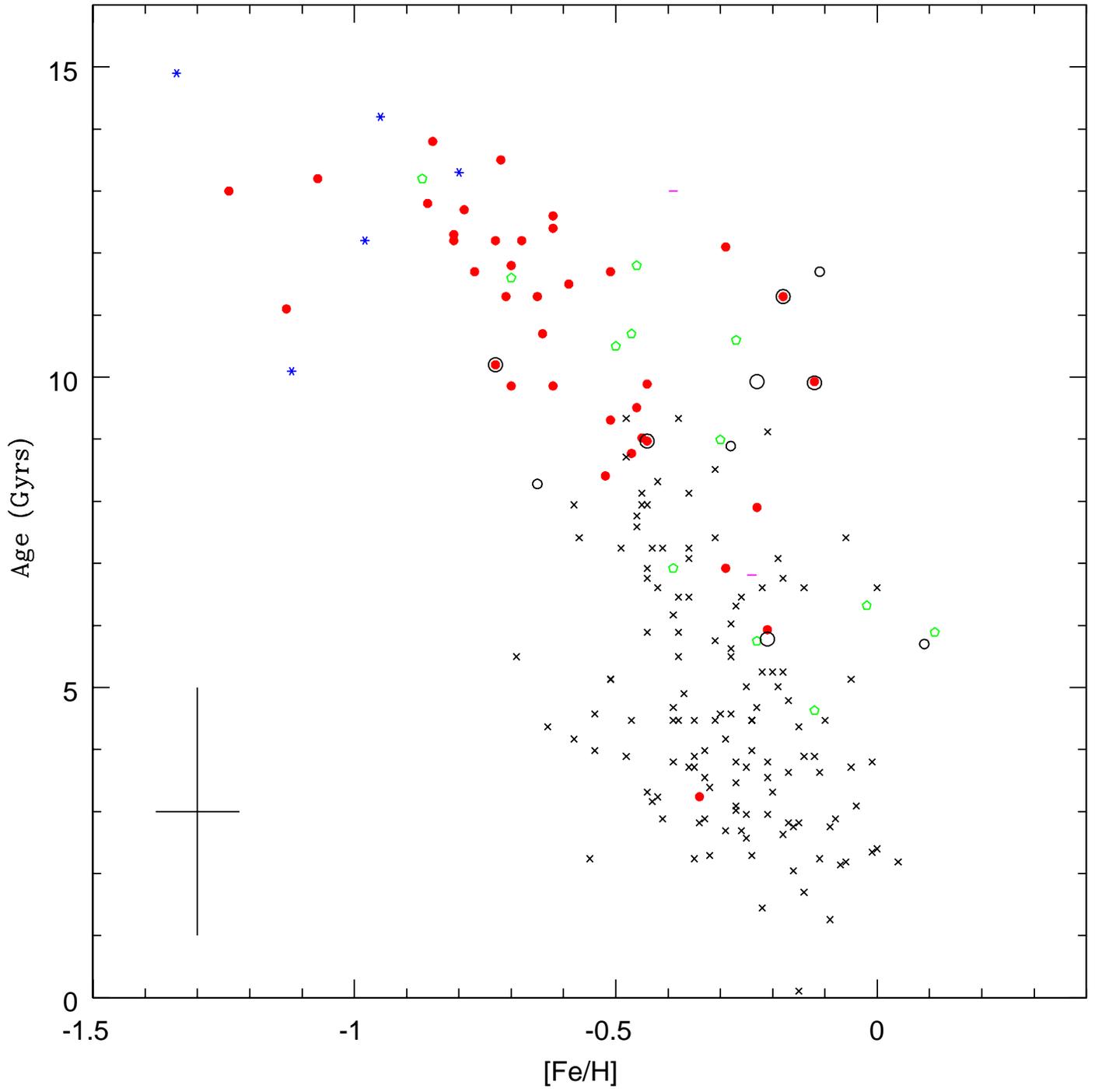
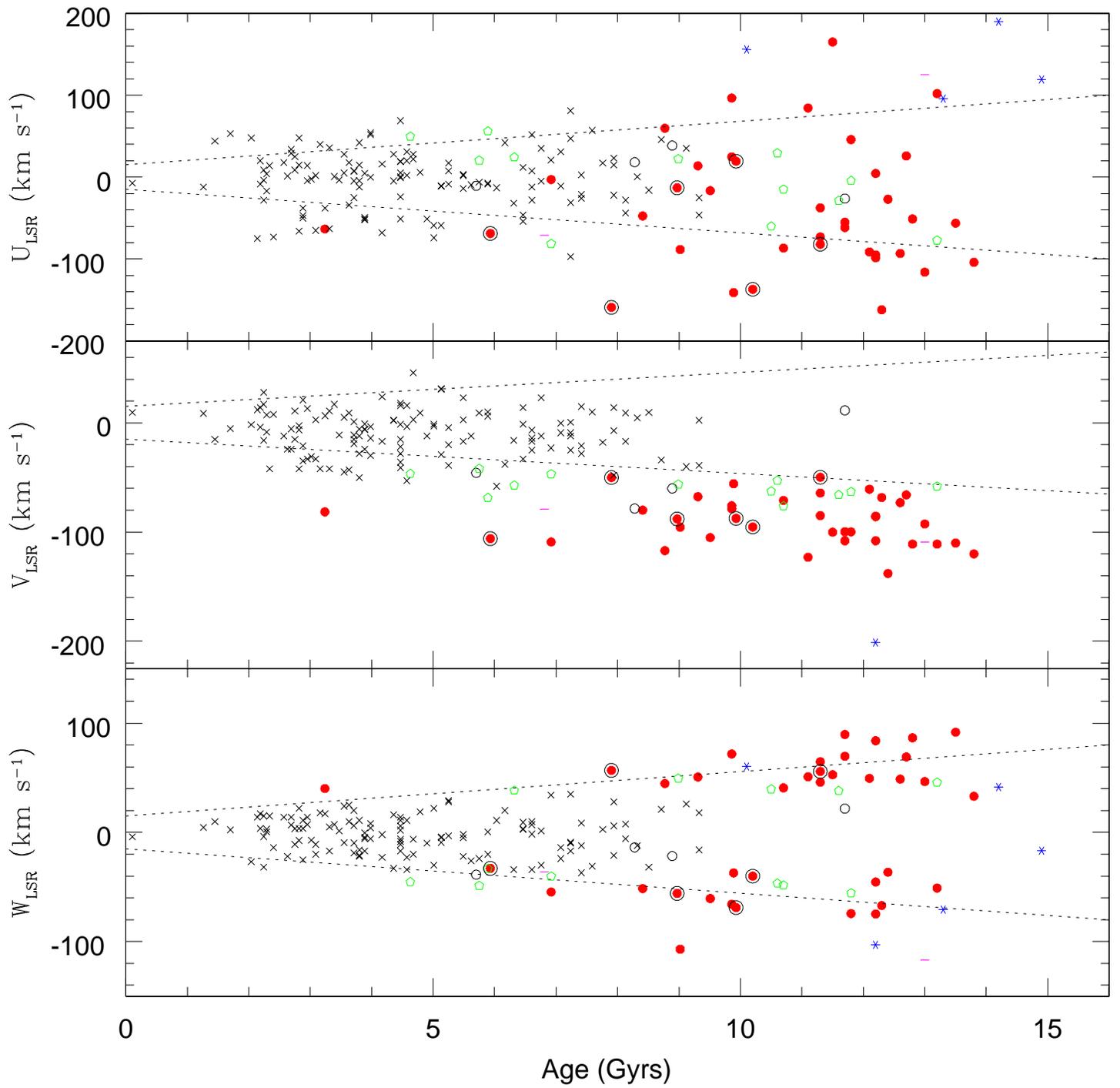
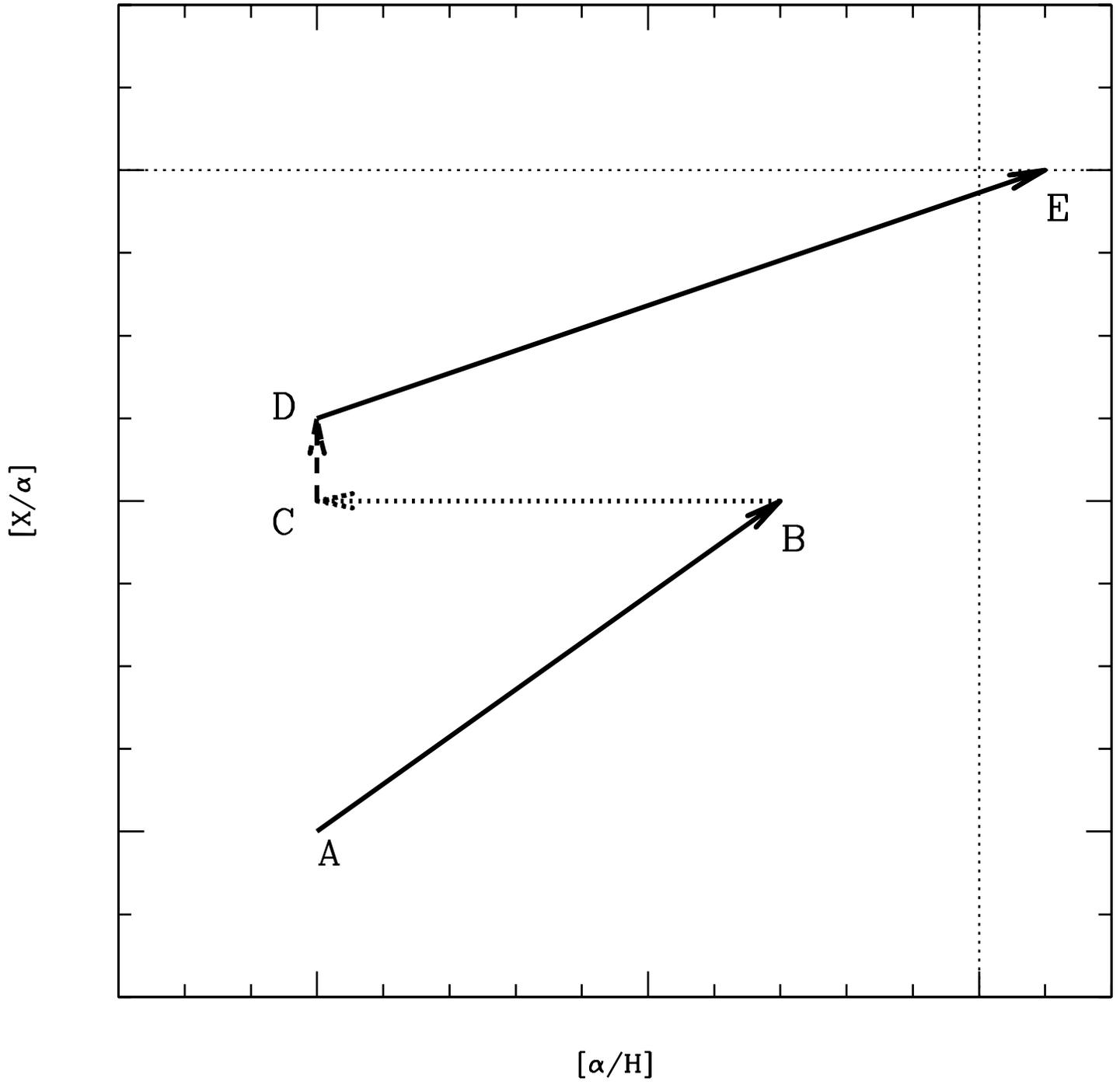
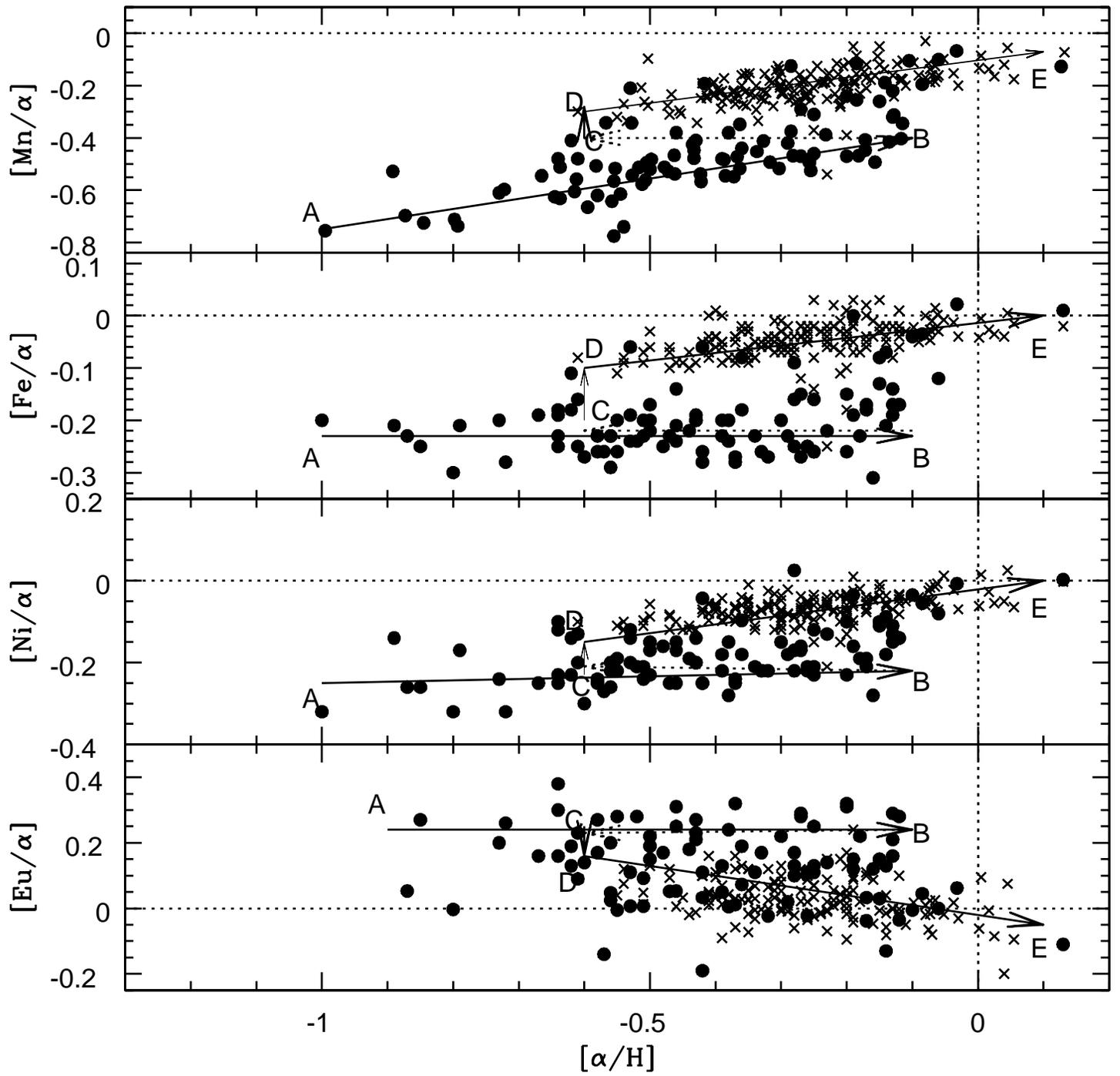
\end{document}